\documentstyle[preprint,eqsecnum,aps,epsf,rotate,floats,graphicx]{revtex}
\newcommand{\bff}[1]{\mbox{\boldmath ${#1}$}}
 
\begin{document}


\renewcommand{\thefootnote}{\fnsymbol{footnote}}
\begin{titlepage}
\begin{flushright}
CERN-TH/99-356\\
\mbox{TTP99-55\hspace*{0.6cm}}\\
hep-ph/0001062\\
\mbox{December 1999\hspace*{0.25cm}}\\
\end{flushright}
\vskip 0.7cm
\begin{center}
 \boldmath
{\Large\bf Quarkonium momentum distributions\\[0.2cm]
in photoproduction and $B$ decay}
\unboldmath
\vskip 1.3cm
{\sc M. Beneke} \hspace{0.2cm}and \hspace{0.2cm} 
{\sc G.A. Schuler}
\vskip .3cm
{\it Theory Division, CERN, CH-1211 Geneva 23, Switzerland}
\vskip .4cm
and
\vskip .4cm
{\sc S. Wolf}
\vskip .3cm
{\em Institut f\"ur Theoretische Teilchenphysik, Universit\"at Karlsruhe,\\
D-76128 Karlsruhe, Germany}
\vskip -0.5cm
\end{center}

\begin{abstract}
\noindent\hspace*{-0.34cm} 
According to our present understanding many $J/\psi$ production 
processes proceed through a coloured $c\bar{c}$ state followed by 
the emission of soft particles in the quarkonium rest frame. The 
kinematic effect of soft particle emission is usually a higher-order 
effect in the non-relativistic expansion, but becomes important near 
the kinematic endpoint of quarkonium energy (momentum) distributions. 
In an intermediate region a systematic resummation of the 
non-relativistic expansion leads to the introduction of so-called 
`shape functions'. In this paper we provide an implementation of 
the kinematic effect of soft gluon emission which is consistent 
with the non-relativistic shape function formalism in the region 
where it is applicable and which models the extreme endpoint region. We 
then apply the model to photoproduction of $J/\psi$ and $J/\psi$ 
production in $B$ meson decay. A satisfactory description of $B$ decay 
data is obtained. For inelastic charmonium photoproduction we conclude 
that a sensible comparison of theory with data requires a transverse 
momentum cut larger than the currently used 1 GeV.\\  

\end{abstract}

\vfill

\end{titlepage}

\section{Introduction}

\noindent
Inclusive charmonium production processes can be 
expressed in a factorized form, combining a short-distance expansion 
with the use of a non-relativistic QCD Lagrangian (NRQCD) \cite{BBL95}. 
The short-distance expansion works 
best for total production cross sections, provided the expansion 
parameter $v^2$ (of order of the typical velocity squared of the quarks in the 
bound state) is small enough. It follows that contrary to 
prior belief many charmonium production processes such as 
production in hadron-hadron collisions at large transverse momentum 
\cite{BF95} 
and at fixed target \cite{BR96}, and in $B$ meson decay 
\cite{BBYL92,KLS96a,KLS96b,BMR99}, are actually dominated 
by production of a coloured $c\bar{c}$ state, followed by a long-distance 
transition to charmonium and light hadrons \cite{revs}.

The theoretical prediction of charmonium energy distributions is more 
delicate. A long-standing problem for the NRQCD factorization approach 
concerns the $z$-distribution in inelastic $J/\psi$ photoproduction, 
where $z=E_{J/\psi}/E_\gamma$ is the quarkonium energy fraction in the 
proton rest frame. The colour 
octet contributions to this quantity grow rapidly near $z=1$ 
\cite{CK96,KLS96b}, 
in conflict with observation \cite{PHOTO}, unless the NRQCD 
matrix elements that normalize the colour octet contribution are made 
rather small.\footnote{There may be other difficulties for the NRQCD 
factorization approach, which we do not discuss in this paper. For 
a long time, transverse polarization of $J/\psi$ produced in hadron-hadron 
collisions at large transverse momentum \cite{CW95,BK97,Lei97} 
has been regarded as 
the crucial test of the theoretical framework. If recent indications 
from CDF of no polarization \cite{CDF} are confirmed by higher statistics 
data, this may indicate a problem with factorization, as suggested in 
\cite{HP98}, or it may imply large spin-symmetry violating corrections.}

One of the physical origins of this discrepancy is as follows: 
the fragmentation  
of the coloured $c\bar{c}$ state into $J/\psi$ occurs via the emission 
of gluons with small momentum fractions of order $v^2$. Because the 
momentum of these gluons is small compared to the momenta involved in the 
hard subprocess that creates the $c\bar{c}$ state, it is neglected in 
leading order in the short-distance expansion (in $v^2$); the 
fragmentation into $J/\psi$ is described by a single number (the `NRQCD 
matrix element'). This is adequate for total production cross sections, 
but it is not for distributions in the kinematic region, where the charmonium 
carries nearly maximal energy. In this region, the $J/\psi$ energy 
distribution is evidently sensitive to the energy distribution of the 
soft emitted gluons. In particular, we expect that the $J/\psi$ energy 
distribution should fall to zero, rather than grow, near the point of 
maximal energy, if the $J/\psi$ is produced via a colour octet state, 
since the emission of gluons with momentum much smaller than their typical 
one is rather unlikely.

The inadequacy of a leading-order treatment of the short-distance 
expansion, and the necessity to account for the kinematics of soft 
gluon emission, is even more evident for $J/\psi$ production in $B$ 
meson decay. The leading order partonic short-distance process 
$b\to (c\bar{c})q$ results in $c\bar{c}$ pairs with fixed (maximal) 
energy, in stark contrast to the broad energy distribution observed 
\cite{CLEO95}. The broad energy distribution of multi-body final 
states has to be attributed to soft gluon emission and to the Fermi 
motion of the $b$ quark in the $B$ meson.

Technically speaking, the velocity expansion of the short-distance 
process breaks down near the kinematic endpoint of maximal charmonium 
energy \cite{MW97,BRW97}, because higher-order terms in the small 
parameter $v^2$ are compensated by inverse powers of small kinematic 
invariants. Such a breakdown of the short-distance expansion 
is not specific to quarkonium production in the NRQCD approach, but 
occurs quite generally for inclusive processes, 
for example in deep-inelastic scattering as Bjorken 
$x\to 1$ or in semi-leptonic 
or radiative $B$ decays \cite{N94}. When the quarkonium 
carries a fraction $(1-\epsilon)$ of its maximal energy, 
where $\epsilon$ is small, the 
inclusiveness of the process is restricted by the small phase-space left 
for the emission of further particles. The process is 
then also sensitive to the fact that the physical phase space is limited by 
hadron kinematics, while the calculation of short-distance coefficients 
is carried out in terms of partons. The short-distance expansion 
reacts to this non-inclusiveness by exhibiting terms of order 
$(v^2/\epsilon)^k$. In some cases one can sum the leading terms in 
$v^2/\epsilon$ to all orders and express the quarkonium production cross 
section as a convolution of a non-perturbative `shape function' with 
a partonic cross section. The shape function leads to a smearing of 
the energy spectrum. The shape function formalism is analogous to 
a leading twist approximation, and is appropriate for 
$\epsilon \sim v^2$. In this intermediate region the framework of the 
NRQCD factorization approach is still valid, reorganized by a partial 
resummation of the velocity expansion. However, 
in the extreme endpoint region, 
$\epsilon\ll v^2$, the twist expansion also breaks down.

The leading twist expressions for several energy 
distributions have been derived in \cite{BRW97}. 
But since the shape function is 
non-perturbative and essentially unknown, no quantitative analysis 
has been performed. It is the aim of this paper to explore the kinematic 
effect of soft gluons in the fragmentation of a coloured $c\bar{c}$ pair 
quantitatively. In particular, we will be interested in the question 
whether folding the short-distance cross section with a shape function 
can indeed account for the observed $z$-distribution in $J/\psi$ 
photoproduction. The emission of soft gluons with energy of order $m_c v^2$ 
in the quarkonium rest frame cannot be computed perturbatively and 
we have to model it. Our ansatz for the soft gluon radiation 
function will be guided 
by simplicity. The important feature 
of the model is that it incorporates the kinematics of soft gluon 
radiation together with reasonable assumptions on the typical energy 
scales involved. The ansatz bears some similarities with Fermi motion 
smearing \cite{AP79} and, in particular, the ACCMM model \cite{ACM} for 
semileptonic $B$ decays. Since the precise form of the energy 
distribution near the endpoint depends on the ansatz for the shape 
function, our results do not constitute theoretical predictions. 
However, as we shall see, a satisfactory description of $B$ decay  
data can indeed be obtained with a reasonable ansatz for 
the shape function. A further cross check is provided by applying the 
same shape function to the $J/\psi$ energy distribution in 
photoproduction. This however, turns out to be more problematic.

The paper is divided as follows: Sect.~2 is `theoretical'. 
We define the model and derive the equation that describes the 
convolution of the short-distance process with the shape function for 
a general production process. We also show that the model is  
equivalent to a specific form of the NRQCD shape function in the region 
where a leading twist approximation is valid. To illustrate the formalism, 
we consider the limit $m_c v^2\gg \Lambda_{\rm QCD}$, in which charmonium 
is a Coulomb bound state. We rederive NRQCD factorization for this 
specific case and compute the shape function in this limit.

The ansatz for the non-perturbative shape function depends on a few 
model parameters. In Sect.~3 we apply the model to 
the $J/\psi$ momentum distribution in $B\to J/\psi+X$ and tune the 
parameters of the model to the observed momentum distribution. 
In Sect.~4 the more complicated (and more interesting) case of 
inelastic photoproduction is considered. 

\section{Shape function model}

\noindent
In this section we derive the general expression for the smeared 
quarkonium energy distributions on which the applications to 
$B$ decay and photoproduction will be based. To motivate our approach 
and to make more explicit contact with the formalism of 
\cite{BBL95,BRW97}, we begin by considering the production amplitude 
for quarkonium in the Coulomb limit, and with emission of a single 
soft gluon, before generalizing the expressions to the case of 
interest. In the last subsection we return to the Coulomb limit 
and compute the shape function in this limit. This provides us with  
an idea of the form of the shape function in a controlled, although 
unrealistic limit.

\subsection{Factorization and the shape function in the 
Coulomb limit}

\begin{figure}[t]
   \vspace{-3cm}
   \epsfysize=29cm
   \epsfxsize=20cm
   \centerline{\epsffile{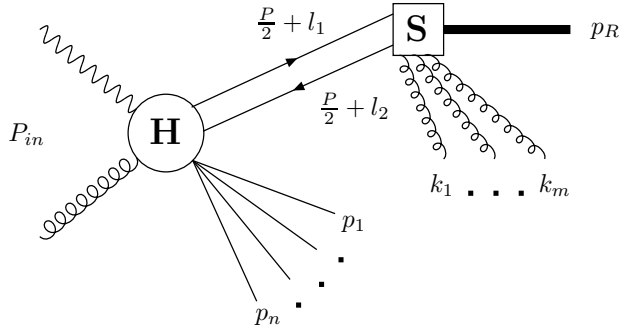}}
   \vspace*{-21.0cm}
\caption[dummy]{\label{fig1}\small 
Diagrammatic representation of the amplitude that leads to 
Eq.~(\ref{start}).}
\end{figure}

\noindent 
Inclusive charmonium production proceeds in two stages \cite{BBL95}: 
first a pair of nearly on-shell and co-moving charm quarks is created 
in a hard process in which typical momenta are of order $2 m_c$ 
(or larger, if there is another hard scale) in the charmonium rest 
frame. The nearly on-shell $c\bar{c}$ state then fragments into 
charmonium via emission of soft particles with energy and momentum of 
order $m_c v^2$ in the charmonium rest frame.\footnote{The energy scale for 
these particles is set by the small velocity $v$ that characterizes 
the non-relativistic charmonium bound state and the typical virtuality 
$(m_c v)^2$ of the nearly on-shell $c$ and $\bar{c}$ quark. See also 
the discussion below.} Schematically, the differential cross section 
is expressed in the factorized form 
\begin{eqnarray}
\label{start}
(2\pi)^3\,2p_R^0\frac{d\sigma}{d^3p_R} &=& \mbox{flux}\,\cdot\int
d\mbox{PS}[p_i,k_j]\,(2\pi)^4\delta(p_R+\sum_j k_j+\sum_i p_i-
P_{in})
\nonumber\\
&&\hspace*{-1.5cm}\cdot\,
H(P_{in},P,l_1,l_2,p_i)\,S(p_R,P,l_1,l_2,k_j),
\end{eqnarray}
where $d\mbox{PS}[p_i,k_j]$ denotes the phase space measure for the 
sets of hard ($p_i$) and soft ($k_j$) particle momenta and $H$ and 
$S$ refer to the hard and soft parts of the amplitude squared, 
respectively. See Fig.~\ref{fig1} for a graphical representation 
and further explanation of notation.\footnote{In an abuse of 
notation, in the figure $H$ and $S$ refer to the hard and soft part of the 
amplitude, rather than the amplitude squared. The nearly on-shell 
heavy quark propagators that connect the hard and soft part in the figure 
should be considered as part of $S$. See below.}

To define the hard and soft parts in (\ref{start}) accurately, we use the 
amplitude for the process $\gamma g\to J/\psi g g$, relevant to inelastic 
photoproduction, as an example. It is also instructive to take the 
limit $m_c v^2\gg \Lambda_{\rm QCD}$, where $v$ is now of order 
$\alpha_s(m_c v)$ and $\Lambda_{\rm QCD}$ is the strong interaction scale. 
We call this the Coulomb limit, because the charmonium 
bound state is perturbatively calculable in this limit 
and the dominant binding is 
through the Coulomb force. The Coulomb limit is much stronger than the 
non-relativistic limit. While charmonium and bottomonium are 
non-relativistic ($v^2\ll 1$), they are not Coulombic ($m_c v^2\sim 
\Lambda_{\rm QCD}$) in reality. In particular, the NRQCD matrix elements, 
which are usually taken as non-perturbative parameters, can be 
perturbatively calculated in the Coulomb limit, up to corrections 
suppressed by powers of $\Lambda_{\rm QCD}/(m_c v^2)$. 

\begin{figure}[t]
   \vspace{-3.5cm}
   \epsfysize=30cm
   \epsfxsize=20cm
   \centerline{\epsffile{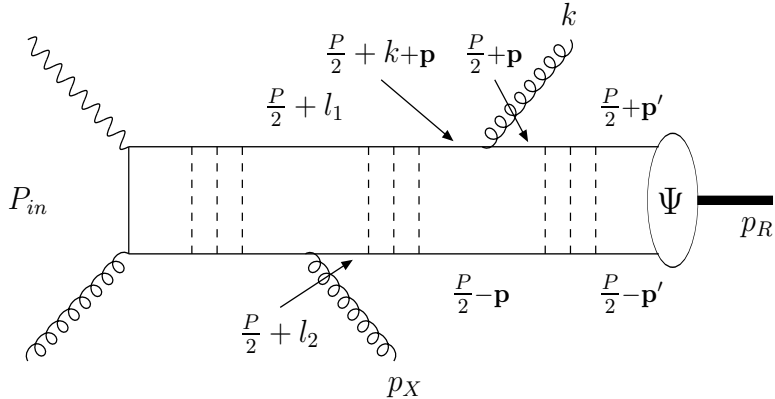}}
   \vspace*{-20.7cm}
\caption[dummy]{\label{fig2}\small 
A contribution to the $\gamma g\to J/\psi g g$ amplitude discussed in the 
text.}
\end{figure}

A particular contribution to the $\gamma g\to J/\psi g g$ amplitude 
is shown in Fig.~\ref{fig2}. The corresponding squared amplitude is the 
sum of terms where both gluons are hard or both gluons are soft or 
one of them is hard and the other is soft. The hard-soft term is the 
most interesting one for inelastic photoproduction through the colour 
octet mechanism and we focus on it first. The other two terms will be 
briefly discussed later.

Suppose the gluon with momentum $p_X$ in Fig.~\ref{fig2} is hard and 
the gluon with momentum $k$ is soft. On-shell soft gluons in NRQCD can 
have energy of order $m_c v$ and $m_c v^2$ \cite{BS98} (called `soft' and 
`ultrasoft', respectively, in \cite{BS98}). However, gluons with 
energy of order $m_c v$ cannot be radiated over the time-scale 
$1/(m_c v^2)$ and do not appear as final state particles in the scattering 
amplitude.\footnote{More technically, because the interaction with a 
gluon with energy of order $m_c v$ sends the heavy quark propagator 
off-shell, a subgraph with energy and momentum 
of order $m_c v$ in the amplitude squared has no $c\bar{c}+n g$ cut, 
as would be required for a non-zero 
contribution to the $\gamma g\to J/\psi g g$ amplitude. Rather such a 
subgraph can be expanded into a 
series of instantaneous interactions, which contribute to the potential 
between the heavy quarks.} Consequently, the scale of $k$ is $m_c v^2$. 
This is important, because this will set the scale for the energy of 
soft gluon emission in our model parametrization later.

In Fig.~\ref{fig2} we included (dashed lines) 
the instantaneous exchange of (Coulomb) 
gluons with energy of order $m_c v^2$ and momentum of order $m_c v$. If 
this exchange occurs between nearly on-shell heavy quark propagators 
with off-shellness of order $m_c v^2$, it is not suppressed by the 
small coupling constant, because the total contribution from each 
gluon is of order $\alpha_s(m_c v)/v\sim 1$. However, if one of the heavy 
quark propagators is far off-shell, Coulomb exchange represents an 
ordinary higher-order correction to the amplitude. Hence we can neglect 
gluon exchange to the left of the gluon with momentum $p_X$. The gluon 
ladder `between' the emission of the gluons with momentum $p_X$ and $k$, 
respectively, cannot be neglected, but it is summed into the 
Coulomb Green function $G_c(\bff{p},\bff{p}';E)$, the Green function 
for the Schr\"odinger equation with the (leading order) Coulomb 
potential. The Green function is related to the quark-antiquark 
scattering amplitude for a quark-antiquark pair with (small) relative 
three momentum $2\bff{p}$ into a quark-antiquark pair with (small) relative 
three momentum $2\bff{p}'$ with total non-relativistic 
energy $E$. Likewise the gluon 
ladder to the right of the gluon with momentum $k$ is summed and  
contained in the bound state wave function. For a ${}^3\!S_1$ state, 
such as $J/\psi$, the bound state wave function in the bound state 
rest frame is given by 
\begin{equation}
\Psi(p_R,\lambda;\bff{p}) = \sqrt{2 M_R}\cdot\frac{\delta_{ab}}{\sqrt{N_c}}
\cdot\frac{\epsilon^i(\lambda) \sigma^i_{\alpha\beta}}{\sqrt{2}}\cdot 
\psi(\bff{p}),
\end{equation}
where
\begin{equation}
\label{wavefn}
\psi(\bff{p}) = \frac{8\sqrt{\pi}\gamma^{5/2}}{(\bff{p}^2+\gamma^2)^2}
\end{equation}
and $\gamma=m_c C_F\alpha_s/2$. $M_R$ is the quarkonium mass, 
$\delta_{ab}$ refers to colour (with $N_c=3$ the number of colours and 
$C_F=(N_c^2-1)/(2 N_c)$), 
$\sigma^i$ is a Pauli matrix and $\epsilon^i(\lambda)$ the polarization 
vector of the quarkonium.

With these remarks one of the two (symmetric) hard-soft contributions 
to the amplitude can be written as 
\begin{eqnarray}
\label{ampl}
{\cal A}(\gamma g\to J/\psi gg) &=& \int \!\frac{d^3\bff{q}}{(2\pi)^3}
\frac{d^3\bff{p}}{(2\pi)^3} \,\,
\hat{{\cal A}}(\gamma g\to c\bar{c}g)\cdot i G_c(\bff{q},
\bff{p}+\bff{k}/2;E(p_R+k))
\nonumber\\
&&\hspace*{-1.5cm}
\cdot \,V(k;\bff{p})\cdot 
\Psi(p_R,\lambda;\bff{p}),
\end{eqnarray}
where $V(k;\bff{p})$ refers to the vertex at which the soft gluon 
is emitted and $\hat{{\cal A}}(\gamma g\to c\bar{c}g)$
denotes the hard sub-amplitude with the on-shell 
spinors for its external heavy quark lines with momentum $P/2+l_1$ and 
$P/2+l_2$ removed. We also introduced the vector $P$, defined 
as $P=(2 m_c,\bff{0})$ in the $J/\psi$ rest frame, the relative momentum 
$q=(l_1-l_2)/2$ and $E(p_R+k)=p_R^0+k^0-2 m_c=-m_c (C_F\alpha_s)^2/4+k^0$. 
The binding energy at leading order has to be kept in the last 
expression, because it is of the same order as $k^0$. For later use 
we define $l=l_1+l_2$, the vector that describes the motion of the 
$c\bar{c}$ pair in the $J/\psi$ rest frame. Note the kinematic relation 
$P+l=p_R+k$.

The amplitude is not yet in a factorized form, because the hard 
sub-amplitude still depends on $q$ and $l$ and its spin and colour 
indices are entangled with those of the remaining part of the 
amplitude. As described in \cite{BBL95}, we can perform a spin 
and colour decomposition that disentangles the two parts of the 
amplitude. We then expand the hard sub-amplitude in the small momentum  
$q$, which amounts to an expansion in derivative operators and 
a decomposition in orbital angular momentum. As a matter of principle, 
we could also expand the hard sub-amplitude in $l$. However, since it is 
$l$ that occurs in the phase space constraint and that is related to 
the terms in the short-distance expansion, which we intend to sum 
to all orders, we do not perform this expansion. The spin and colour 
decomposition, and the expansion in relative momentum $q$, results in 
the following expansion of the amplitude squared:
\begin{eqnarray}
\label{ff1}
\left|{\cal A}(\gamma g\to J/\psi gg)\right|^2 &=& 
\sum_n \,\mbox{Pr}_n\left[\hat{{\cal A}}(\gamma g\to c\bar{c}g)\right]
\,\mbox{Pr}_n^\prime\left[\hat{{\cal A}}(\gamma g\to c\bar{c}g)^\star\right]
\nonumber\\
&&\hspace*{-3.0cm}
\cdot \int \!\frac{d^3\bff{q}}{(2\pi)^3}\frac{d^3\bff{p}}{(2\pi)^3} \,
\mbox{tr}\left[\Gamma_n(\bff{q})\,i G_c(\bff{q},
\bff{p}+\bff{k}/2;E(p_R+k))
\,V(k;\bff{p})\,\Psi(p_R,\lambda;\bff{p})\right]
\nonumber\\
&&\hspace*{-3.0cm}
\cdot 
\int \!\frac{d^3\hat{\bff{q}}}{(2\pi)^3}\frac{d^3\hat{\bff{p}}}{(2\pi)^3} \,
\mbox{tr}\left[\Gamma_n^\prime(\hat{\bff{q}})\,i G_c(\hat{\bff{q}},
\hat{\bff{p}}+\bff{k}/2;E(p_R+k))
\,V(k;\hat{\bff{p}})\,
\Psi(p_R,\lambda;\hat{\bff{p}})\right]^\star.
\end{eqnarray}
Here $\Gamma_n(\bff{q})$ is a matrix in spin and colour indices and a 
polynomial in $\bff{q}$. The operator $\mbox{Pr}_n$ is also a matrix 
in spinor and colour indices and extracts the appropriate 
Taylor coefficient of expansion of $\hat{{\cal A}}(\gamma g\to c\bar{c}g)$
in $\bff{q}$. The quantity 
$\mbox{Pr}_n\left[\hat{{\cal A}}(\gamma g\to c\bar{c}g)\right]$ 
is $\bff{q}$-independent, but still depends on $l$. In conventional 
NRQCD terms, the sum over $n$ corresponds to intermediate $c\bar{c}$ pairs 
in different angular momentum and colour states, and also to higher 
dimension operators in each intermediate channel.
The previous equation can be written as the product of a hard and soft part,
\begin{equation}
\left|{\cal A}(\gamma g\to J/\psi gg)\right|^2 = 
\sum_n\,H_n(P_{in},P,l,p_X)\,S_n(p_R,P,k),
\end{equation}
where the soft part $S_n$ is given by the last two lines of (\ref{ff1}). 
$H_n$ and $S_n$ are still coupled through the relation $P+l=p_R+k$, 
so we introduce $1=\int d^4l\,\delta(p_R+k-P-l)$. Adding the 
phase space integration over $p_X$ and $k$, we recover the 
differential cross section in a form similar to (\ref{start}):
\begin{eqnarray}
\label{ff2}
(2\pi)^3\,2p_R^0\frac{d\sigma}{d^3p_R} &\equiv&
\sum_n\int \!\!\frac{d^4 l}{(2\pi)^4}\,\,
\hat{\sigma}(c\bar{c}[n])(l)\cdot F_n(l) 
\nonumber\\
&&\hspace*{-2.0cm}
=\,\sum_n\int \!\!\frac{d^4 l}{(2\pi)^4}\,\,
\mbox{flux}\,\int d\mbox{PS}[p_X]\,(2\pi)^4\delta(P+l+p_X-P_{in})\,
H_n(P_{in},P,l,p_X)\nonumber\\
&&\hspace*{-1.5cm}\,\cdot\int
d\mbox{PS}[k]\,(2\pi)^4\delta(p_R+k-P-l)\,S_n(p_R,P,k),
\end{eqnarray}
where $\hat{\sigma}(c\bar{c}[n])(l)$ refers to the short-distance part 
and $F_n(l)$ to the soft part. The expansion in local operators appropriate 
to integrated cross sections \cite{BBL95} is recovered after expansion 
of $H_n(P_{in},P,l,p_X)$ in $l$. In leading order, we then identify 
\begin{equation}
\label{soft1}
\int d\mbox{PS}[k]\,S_n(p_R,P,k)
\end{equation}
with the NRQCD matrix elements defined in \cite{BBL95}.

Before continuing let us discuss as an example the angular momentum and 
colour projection for the case of an intermediate $c\bar{c}$ pair in a 
${}^3\!S_1$, colour-singlet state, at lowest order in the expansion in 
$\bff{q}$. In this case $\mbox{Pr}$ simply sets $\bff{q}$ to zero in 
the hard sub-amplitude and $\Gamma(\bff{q})$ carries no 
$\bff{q}$-dependence. The correctly normalized spin and colour 
projection is 
\begin{eqnarray}
&&\mbox{Pr}_n \,[\ldots]\,\,\rightarrow\,\, 
\frac{1}{\sqrt{2\!\cdot \!2 m_c}}\cdot \frac{1}{\sqrt{3}}
\cdot \frac{1}{2 N_c}\cdot \mbox{tr}(\not\!\epsilon_\lambda(P) 
(\not\!P+2 m_c) [\ldots]),
\\
\label{proj2}
&&\Gamma_n(\bff{q})\otimes \Gamma_n^\prime(\bff{q})
\,\,\rightarrow \,\,
\frac{1}{\sqrt{2\!\cdot \!2 m_c}}\cdot \sigma^i \otimes 
\frac{1}{\sqrt{2\!\cdot \!2 m_c}}\cdot \sigma^i,
\end{eqnarray}
where the trace includes a colour trace and the projection of the hard 
amplitude is written in a covariant form. Let us check that (\ref{soft1}) 
together with the projection (\ref{proj2}) 
do indeed reproduce the colour singlet NRQCD matrix element. 
In leading order the transition ${}^3\!S_1^{(1)} \to {}^3\!S_1^{(1)}$ 
does not require gluon emission. Hence 
\begin{eqnarray}
&&\int d\mbox{PS}[k]\,S_n(p_R,P,k) \,\,\rightarrow \,\,
\nonumber\\
&&\hspace*{1.5cm}
\sum_\lambda
\left|\,\int \!\frac{d^3\bff{p}}{(2\pi)^3} \,\frac{\mbox{tr}(\sigma^i
\Psi(p_R,\lambda;\bff{p}))}{\sqrt{2\!\cdot \!2 m_c}}\right|^2 
=\frac{M_R}{2 m_c}\cdot 6N_c|\Psi(0)|^2 \approx 
\langle {\cal O}_1({}^3\!S_1)\rangle,
\end{eqnarray}
where we used that in the leading order approximation 
$M_R\approx 2 m_c$.

Note that $F_n(l)$ in (\ref{ff2}) 
defines a more general object than the shape function 
in \cite{BRW97}, which is a function of only one variable $l_+=l_0+l_z$ 
or $l_0$. The definitions of \cite{BRW97} would be reproduced, if we 
could neglect the other components of $l$ in the short-distance part. 
We shall discuss later, after generalizing (\ref{ff2}) to the emission 
of more than one gluon, under what conditions this is justified.

Up to now we considered the contribution of the diagram in 
Fig.~\ref{fig2} to $J/\psi$ photoproduction, when one of the two emitted 
gluons is hard and the other is soft. The contribution from 
two hard gluons is part of the next-to-leading order correction 
to the short-distance part of the colour-singlet intermediate state.
The contribution from two soft gluons smears out the contribution from 
the diagram with no gluon emission, which is concentrated at $z=1$ and 
zero transverse momentum. It also contributes to the endpoint 
of the energy spectrum, but can be eliminated with a transverse momentum 
cut sufficiently large compared to several hundred MeV. Experimental 
measurements of inelastic $J/\psi$ photoproduction usually imply 
such a cut.

\subsection{The general case}

\noindent
We now extend the previous discussion in the following way. We consider 
a general, inclusive charmonium production process (cf.~Fig.~\ref{fig1})
\begin{equation}
\mbox{Initial state\,} (P_{in}) \to c\bar{c}[n] + X(p_i) \to 
J/\psi(p_R) +X(p_i) + Y(k_j),
\end{equation}
where the $c\bar{c}$ pair is in a certain colour and angular momentum 
state $n$, $X$ denotes a collection of hard particles, and $Y$ a collection 
of soft particles emitted in the fragmentation of the $c\bar{c}$ pair.

Since $m_c v^2\sim \Lambda_{\rm QCD}$, the coupling to soft gluons 
is large and the emission of multiple gluons is not suppressed. Hence 
the emission of soft gluons is better described as the emission of a 
soft colour multipole field, which carries away a total momentum 
$k=\sum_j k_j$ and which has the correct quantum numbers to effect 
the transition from $c\bar{c}[n]$ to $J/\psi$. Hence 
we define
\begin{equation}
\label{phi}
\Phi_n(k;p_R,P) = \int d\mbox{PS}[k_j]\,
(2\pi)^4\delta(k-\sum_j k_j)\,S_n(p_R,P,k_j),
\end{equation}
where $S_n(p_R,P,k_j)$ is the generalization of the soft sub-amplitude 
that appears in (\ref{ff2}) to the emission of more than one soft gluon.
With this definition the generalization of (\ref{ff2}) is given by
\begin{eqnarray}
\label{vers1}
(2\pi)^3\,2p_R^0\frac{d\sigma}{d^3p_R} &\equiv&
\sum_n\int \!\!\frac{d^4 l}{(2\pi)^4}\,\,
\hat{\sigma}(c\bar{c}[n])(l)\cdot F_n(l) 
\nonumber\\
&&\hspace*{-2.0cm}
=\,\sum_n\int \!\!\frac{d^4 l}{(2\pi)^4}\,\,
\mbox{flux}\,\int d\mbox{PS}[p_i]\,(2\pi)^4\delta(P+l+\sum_i p_i-P_{in})\,
H_n(P_{in},P,l,p_i)\nonumber\\
&&\hspace*{-1.5cm}\,\cdot
\left[\,\int\frac{d k^2}{2 \pi}\frac{d^3 \bff{k}}{(2\pi)^3 2 k^0}\,
(2\pi)^4\delta(p_R+k-P-l)\,\Phi_n(k;p_R,P)\right]\!.
\end{eqnarray}
As above the differential cross section is factored into a short-distance 
and a soft part. In higher orders in the strong coupling, this would 
require careful subtractions to define both parts properly. We will be 
working only with cases, where the lowest order, tree approximation 
to the short-distance part is assumed. Then the factorization is trivial, 
as in the example of the previous subsection.

There is an additional assumption implicit in writing (\ref{vers1}), 
which concerns the validity of NRQCD factorization in general \cite{BBL95}, 
not only its generalization to spectra. The assumption is that the 
transition from the $c\bar{c}[n]$ state to $J/\psi$ occurs via emission 
of gluons rather than by absorption from the surrounding 
`partonic medium'. Of course, if $n$ is a colour octet state the 
emitted gluons must interact with the remnant process to form colour neutral 
hadrons; the NRQCD approach assumes that the process of colour 
neutralization is suppressed by powers of $\Lambda_{\rm QCD}/m_c$ and 
can be formally ignored, if we consider $v$ and $\Lambda_{\rm QCD}/m_c$ 
as independent parameters such that $\Lambda_{\rm QCD}/m_c\ll v\ll 1$. 
On the other hand, absorption would violate factorization explicitly, 
since its details depend on the environment created by the specific 
production process. Despite the fact that this issue affects most 
quarkonium production processes, it has rarely been addressed in the 
literature, with the exception of \cite{HP98}. We will not dwell on 
this issue further and take factorization for granted. 
(The empirical fact that the NRQCD matrix 
elements are approximately universal, including hadronic collisions, 
may support this assumption.) However, an investigation of this 
point would certainly be useful.

\subsubsection{Derivation of the smeared spectrum}
\label{deriv}

\noindent
We now bring (\ref{vers1}) into a more useful form. We make one additional 
simplification, which is adequate to the two applications which we consider 
in this paper. The simplification is that there is only a single, massless 
hard particle in the final state. Then the set of momenta $p_i$ consists 
of only $p_X$, and $p_X^2=0$. 

It is often convenient to refer explicitly to the quarkonium rest frame 
defined by $\bff{p}_R=\bff{P}=0$ rather then the centre-of-mass frame 
defined by $\bff{P}_{in}=0$. In the following non-invariant quantities 
will refer to the quarkonium rest frame. For example, in $(p_R-P)\cdot l=
(M_R-2 m_c)\,l_0$, $l_0$ refers to the zero-component of $l$ in the 
quarkonium rest frame. We define the $z$-direction as the direction 
of $\bff{P}-\bff{P}_{in}$ in the quarkonium rest frame and in the 
centre-of-mass frame. With  this unconventional definition of the 
$z$-direction in the centre-of-mass frame 
the boost from the centre-of-mass to the 
quarkonium rest frame is in the $z$-direction and the transverse 
components defined with respect to this axis are invariant.

We use the two $\delta$-functions in (\ref{vers1}) to integrate 
over $\bff{p}_X$ and $\bff{k}$. Then define $l_\pm=l_0\pm l_z$ 
and write 
\begin{equation}
\frac{d^4 l}{(2\pi)^4} = \frac{dl_+ dl_0 dl_\perp^2 d\phi}{32\pi^4}.
\end{equation}
The $\delta$-function left over from the second $\delta$-function 
in (\ref{vers1}) fixes
\begin{equation}
\label{lperp}
l_\perp^2 = (M_R-2 m_c)^2-2 (M_R-2 m_c) l_0+l_+ (2 l_0-l_+)-k^2.
\end{equation}
The result of these manipulations is 
\begin{eqnarray}
\label{vers2}
(2\pi)^3\,2p_R^0\frac{d\sigma}{d^3p_R} &=&
\nonumber\\
&&
\hspace*{-1.5cm}\sum_n\int \frac{dk^2}{2\pi}\,dl_+ dl_0\frac{d\phi}{2\pi}\,\,
\delta(A)\cdot \mbox{flux}\cdot
H_n(P_{in},P,l,p_X)\cdot
\frac{1}{4\pi}\Phi_n(k;p_R,P),
\end{eqnarray}
with
\begin{eqnarray}
\label{constraint}
A&\equiv& (2 m_c-P_{in-})\,(2 m_c-P_{in+}+l_+)+(2 l_0-l_+)\,(2 m_c-P_{in+})
\nonumber\\
&&\,-(M_R-2 m_c)\,(M_R-2 m_c-2 l_0)+k^2
\end{eqnarray}
and $p_X=P_{in}-(P+l)$, $k=P+l-p_R$. Furthermore, we have the constraints 
$k_0>0$, $p_{X,0}>0$, $k^2>0$ and $l_\perp^2>0$. 

Any ansatz for the function $\Phi_n(k;p_R,P)$ that we will be using 
will be independent of the azimuthal component $\phi$ of $l$. Hence we need 
only the azimuthally averaged short-distance part:
\begin{equation}
\label{azav}
\bar{H}_n(P_{in},P,l,p_X) \equiv 
\int\frac{d\phi}{2\pi}\, H_n(P_{in},P,l,p_X).
\end{equation}
The remaining $\delta$-function can be used to integrate over $l_+$. Then 
we use $k_0=2 m_c-M_R+l_0$ as integration variable instead of $l_0$ and 
define
\begin{equation}
\label{albe}
\alpha\equiv P_{in+}-M_R,\quad\beta\equiv P_{in-}-M_R.
\end{equation}
This leads to the final result
\begin{eqnarray}
\label{master}
(2\pi)^3\,2p_R^0\frac{d\sigma}{d^3p_R} &=&
\nonumber\\
&&\hspace*{-1.5cm}
\sum_n \int\limits_0^{\alpha\beta} \frac{dk^2}{2\pi}\!
\int\limits_{(\alpha^2+k^2)/(2\alpha)}^{(\beta^2+k^2)/(2\beta)}
\!\!\!dk_0\,\,
\mbox{flux}\cdot\bar{H}_n(P_{in},P,l,p_X)\cdot
\frac{1}{4\pi(\beta-\alpha)}\,\Phi_n(k;p_R,P).
\end{eqnarray}
Recall that $\alpha$, $\beta$ and $k_0$ are defined in the quarkonium 
rest frame.

The integration limits are obtained as follows: inserting the constraint 
(\ref{constraint}) $A=0$ on $l_+$ provided by the last $\delta$-function 
into $l_\perp^2>0$ with $l_\perp^2$ given by (\ref{lperp}), we find the 
condition
\begin{equation}
\label{con2}
\left[k^2-\alpha (2 k_0-\alpha)\right]\,
\left[k^2-\beta (2 k_0-\beta)\right] < 0,
\end{equation}
in addition to $k_0>0$ and $k_0<(\alpha+\beta)/2$, which follows from 
$p_{X,0}>0$. Now note that $\alpha\beta=(P_{in}-p_R)^2>0$ and that 
$k_0>0$ implies $\alpha+\beta>0$. Hence $\alpha$ and $\beta$ are both 
positive. Now $\alpha-\beta=2 P_{in,z}$. In the quarkonium rest frame 
the $z$-axis is defined by the direction of $-\bff{P}_{in}$. This 
implies
\begin{equation}
\beta>\alpha>0.
\end{equation}
Eq.~(\ref{con2}) admits two solutions. The physical one yields the limits 
on the $k_0$-integration in (\ref{master}). The upper limit on the 
$k^2$-integral then follows. Note that $k_0>0$ and $k_0<(\alpha+\beta)/2$ 
are then respected automatically.

Eq.~(\ref{master}) is the main result of this section and we will use it 
later to obtain the $J/\psi$ energy spectra in $B$ decay and 
photoproduction. Recall that $\mbox{flux}\cdot
\bar{H}_n(P_{in},P,l,p_X)$ is just 
the ordinary, projected $c\bar{c}$ production cross section that enters 
familiar applications of NRQCD factorization with the only difference 
that the $c\bar{c}$ pair is produced with momentum $P+l$ rather than 
$P$, and that an average over the azimuthal angle of $l$ in the quarkonium 
rest frame is performed. This means that the invariant mass of the 
$c\bar{c}$ pair is given by $M_{c\bar{c}}^2=(P+l)^2=(p_R+k)^2=
M_R^2+2 M_R k_0+k^2\geq M_R^2$ rather than $4 m_c^2$ as in the conventional 
partonic calculation. This kinematic difference can make a large 
numerical effect.

The radiation function $\Phi_n(k;p_R,P)$ is defined 
by (\ref{phi}). Roughly speaking, it represents the probability squared 
that a soft gluon cluster with energy $k_0$ in the $J/\psi$ rest frame and 
invariant mass $k^2$ is emitted from the $c\bar{c}$ pair in the 
transition $c\bar{c}[n]\to J/\psi$. We consider it as a non-perturbative 
function. We will make an ansatz and try to determine some of its 
parameters from existing data. In the Coulomb limit, the function 
$\Phi_n(k;p_R,P)$ could be computed as indicated previously. However, 
we shall not assume this limit for charmonium.

\subsubsection{The shape function limit}
\label{shapelim}

\noindent
As mentioned above, the function 
\begin{equation}
F_n(l) = \int\frac{d k^2}{2 \pi}\frac{d^3 \bff{k}}{(2\pi)^3 2 k^0}\,
(2\pi)^4\delta(p_R+k-P-l)\,\Phi_n(k;p_R,P)
\end{equation} 
defined in (\ref{vers1}) is different from the shape function 
introduced in \cite{BRW97}. The shape functions introduced there 
correspond to a systematic resummation of enhanced higher order 
corrections in the NRQCD velocity expansion. Eq.~(\ref{master}) goes 
beyond such a systematic resummation. We now show that (\ref{vers1}) 
and (\ref{master}) are equivalent to the results of \cite{BRW97} 
in the region of applicability of the latter, up to non-enhanced 
higher order terms in the velocity expansion.

We are concerned with energy spectra in a variable $z$. For quarkonium 
production in the decay of a heavier particle with mass $m$, we define 
$z=2 P_{in}\cdot p_R/P_{in}^2$. The maximal value of $z$ is 
$z_{max}=1+M_R^2/m^2$, assuming that all other particles in the final 
state are massless. (In reality these will be pions; we neglect 
the small pion mass.) For quarkonium production in two-to-two collisions, 
$a(p_1)+b(p_2)\to J/\psi+\ldots$, we define $z=2 p_2\cdot p_R/P_{in}^2$. 
For example, in $\gamma p$ collisions $p_2$ is the momentum of the struck 
parton in the proton. The maximal value of $z$ is $z_{max}=1$. 

Consider the $z$-spectrum in the region $z_{max}-z$ of order $v^2\ll 1$, 
but $z_{max}-z$ not much smaller than $v^2$. This is the region in which 
the shape function formalism of \cite{BRW97} applies. We introduce 
$p_X=\sum _i p_i$ in (\ref{vers1}) and use the first $\delta$-function 
to integrate over $\bff{p}_X$. This leaves a $\delta$-function 
with argument
\begin{equation}
\label{eq4}
\left[(P-P_{in})_-+l_-\right]\left[(P-P_{in})_++l_+\right] - l_\perp^2-
p_X^2.
\end{equation}
Using the definitions of $z$, it is easy to see that in the endpoint 
region $p_X$ and $P-P_{in}$ become nearly light-like. With our definition 
of the $z$-axis $(P-P_{in})_+$ becomes small, of order $m_c v^2$ 
(but not much smaller), while $(P-P_{in})_-$ remains of order 
$m_c$.\footnote{All other large scales that the process may involve 
are treated as order $m_c$.} $p_X^2$ has to be of order $m_c^2 v^2$ 
or smaller. All components of $l$ scale as $m_c v^2$, since $M_R-2 m_c$ 
and all components of $k$ are of this order. It follows 
that the dependence of (\ref{eq4}) on $l_-$ and $l_\perp$ can 
be dropped. Furthermore, the formalism of \cite{BRW97} assumed 
that the dependence of the hard cross section $H_n(P_{in},P,l,p_X)$ 
on $l$ can be neglected, since it is not related to enhanced higher 
order terms in the velocity expansion. As a consequence, we can pull the 
$l_-$- and $l_\perp$-integrations through to the second line of 
(\ref{vers1}). The result then takes the form of a partonic 
differential production cross section convoluted with a shape function 
in $l_+$, provided we identify the shape function defined in \cite{BRW97} 
with
\begin{eqnarray}
\label{shapefn}
F_n(l_+) &\equiv&
\sum_Y\,\langle 0|\psi^\dagger \Gamma_n\chi|J/\psi+Y\rangle\langle 
J/\psi+Y|\delta(l_+-i D_+)(\chi^\dagger \Gamma_n'\psi)|0\rangle
\nonumber\\
&=& \int\frac{d l_- d^2 l_\perp}{2 (2\pi)^4} 
\left[\,\int\frac{d k^2}{2 \pi}\frac{d^3 \bff{k}}{(2\pi)^3 2 k^0}\,
(2\pi)^4\delta(p_R+k-P-l)\,\Phi_n(k;p_R,P)\right].
\end{eqnarray}
This shows that (\ref{vers1}) is consistent with the operator formalism 
of \cite{BRW97} in the region of $z$ where the operator formalism applies. 
Eqs.~(\ref{vers1}) and (\ref{master}) extrapolate this formalism into the 
extreme endpoint region $z_{max}-z\ll v^2$. Since there is no 
correspondence with a 
systematic resummation of the velocity expansion in the extreme endpoint 
region, this extrapolation should be considered as a model. This is 
again analogous to energy spectra in semileptonic $B$ decays \cite{N94}.

It is instructive to recover the consistency with the shape function 
formalism directly from (\ref{master}). In the region $z_{max}-z\sim v^2$, 
we may approximate (\ref{constraint}) by
\begin{equation}
A\approx (2 m_c-P_{in-})\,(2 m_c-P_{in+}+l_+)=(\alpha+M_R-2 m_c-l_+)
(\beta+M_R-2 m_c).
\end{equation}
This implies that the upper integration limits in (\ref{master}) 
are replaced by infinity.\footnote{This is consistent with 
$\alpha \sim m_c v^2$ and $\beta \sim m_c$ in the shape function 
limit, such that $\alpha\beta\sim m_c^2 v^2$ and $\beta+k^2/\beta \sim 
m_c$, i.e. both upper limits are parametrically larger than the typical 
values of the integration variables $k^2\sim m_c^2 v^4$, and 
$k_0\sim m_c v^2$, respectively.}
We can then re-introduce 
$1=\int d l_+\,\delta(l_+-\alpha-[M_R-2 m_c])$ and 
factorize (\ref{master}) into 
a convolution over the hard cross section times the shape function 
(\ref{shapefn}).

\subsubsection{Form of $\Phi_n(k;p_R,P)$}

\noindent
Eq.~(\ref{shapefn}) implies that the moments of the shape-function are 
related to the usual NRQCD matrix elements. For example, integration 
over $l_+$ results in 
\begin{equation}
\label{normalization}
\int\frac{d^4 l}{(2\pi)^4}\,F_n(l) = 
\frac{1}{(2\pi)^3}\int\limits_0^\infty d k^2\!\!
\int\limits_{\sqrt{k^2}}^\infty \!
d k_0\,\sqrt{k_0^2-k^2}\,\Phi_n(k;p_R,P) = 
\langle {\cal O}^{J/\psi}_n\rangle,
\end{equation}
where $\langle {\cal O}^{J/\psi}_n\rangle$ is the conventional 
NRQCD matrix element for an intermediate $c\bar{c}$ pair in an  
angular momentum and colour state $n$. This could in principle 
be used to determine the overall normalization of $\Phi_n(k;p_R,P)$ 
from the known NRQCD matrix elements.

In practice this is problematic. The phenomenological values of 
the NRQCD matrix elements are determined from integrated quantities 
in leading order in the velocity expansion in a given channel $n$. 
On the other hand, if we compute the same integrated quantities from 
the spectra obtained with (\ref{master}), they contain higher order 
terms in the velocity expansion, for example related to the fact that 
the invariant mass of the $c\bar{c}$ pair is always larger than the 
quarkonium mass $M_R$. Since $v^2$ is not small, the integrated 
quantities can be quite different, if the normalization condition 
(\ref{normalization}) is imposed. Another way of saying this is that 
the phenomenological values of the NRQCD matrix elements would be 
quite different from the commonly accepted ones, if the theoretical 
prediction used to obtain them contained higher order terms in the 
velocity expansion. As a consequence we are forced to tune anew the 
overall normalization to the measured integrated spectra. We will 
return to this point below in the context of specific applications. 

The radiation function $\Phi_n(k;p_R,P)$ is non-perturbative. Similar 
in spirit to the ACCMM model \cite{ACM} for semileptonic $B$ decays, we 
assume a simple functional ansatz for phenomenological studies:
\begin{equation}
\label{ansatz}
\Phi_n(k;p_R,P) = a_n\cdot|\bff{k}|^{b_n} \exp(-k_0^2/\Lambda_n^2)
\cdot k^2\exp(-k^2/\Lambda_n^2).
\end{equation}
The exponential cut-off reflects our expectation that the typical 
energy and invariant mass of the radiated system is of order 
$\Lambda_n\sim m_c v^2\approx \,\,$several hundred MeV. Since the 
pattern of soft gluon radiation may depend on the $c\bar{c}$ state 
$n$, the parameters $a_n$, $b_n$ and $\Lambda_n$ can differ for 
different states. The three parameters of the ansatz could be 
determined from the first three moments of the shape function. 
In practice this is not possible, not only because of the problem 
mentioned above, but also because the NRQCD matrix elements with 
derivatives to which the higher moments are related are not 
known phenomenologically.

In later applications, we will need the radiation functions for 
the three colour octet states $n={}^1\!S_0^{(8)}, {}^3\!P_0^{(8)}, 
{}^3\!S_1^{(8)}$. We assume that
\begin{eqnarray}
\label{bs}
&& b[{}^1\!S_0^{(8)}]=2,\quad b[{}^3\!P_0^{(8)}]=b[{}^3\!S_1^{(8)}]=0,
\\
\label{cutparam}
&& \Lambda[{}^1\!S_0^{(8)}]=\Lambda[{}^3\!P_0^{(8)}]\equiv \Lambda,\quad 
\Lambda[{}^3\!S_1^{(8)}]=c\Lambda. 
\end{eqnarray}
The choice of $b[{}^1\!S_0^{(8)}]=2$ is motivated by the fact that 
the gluon coupling for a M1 magnetic dipole transition from a 
${}^1\!S_0^{(8)}$ to $J/\psi$ is proportional to the momentum of 
the gluon. Furthermore, the transition from  ${}^3\!S_1^{(8)}$ to 
$J/\psi$ occurs through two E1 electric dipole transition, which suggests 
that the average radiated energy and invariant mass is larger than 
for the single M1 and E1 transition in the other two cases. We fix 
$c=1.5$; the effect of this somewhat arbitrary choice will be discussed 
in the context of specific applications. Of course, since soft 
gluon emission is non-perturbative for charmonium, the arguments that 
lead to these choices are at best indicative in any case. 

\subsection{Computation of the shape function in the Coulomb limit}

\noindent
In the following we compute the radiation function in the Coulomb 
limit $m_c v^2 \gg \Lambda_{\rm QCD}$, $\alpha_s(m_c v)\sim v$ for 
$n={}^1\!S_0^{(8)}, {}^3\!P_0^{(8)}$ to obtain an idea of 
the form of this function in a controlled limit. 
Since this limit is unrealistic 
for $J/\psi$, the reader interested only in the application of the formalism 
presented above may jump directly to the next 
section.\footnote{The calculation is similar to a calculation 
reported in \cite{Wong}. However, in this work the $c\bar{c}$ pair 
in state $n$ is described by a Coulomb wave function just as $J/\psi$. 
This substitution does not correspond to the NRQCD definition of 
a colour octet operator or the corresponding shape function, in 
which the $c\bar{c}$ pair is local and all intermediate states 
with the quantum numbers $n$ are allowed, and described by the 
full Coulomb Green function.}

We begin with the chromo-magnetic dipole transition 
$c\bar{c}[{}^1\!S_0^{(8)}]\to J/\psi+g$. With emission of one gluon 
(\ref{phi}) simplifies to 
\begin{equation}
\Phi[{}^1\!S_0^{(8)}](k;p_R,P) = 
2\pi\delta(k^2)\,S[{}^1\!S_0^{(8)}](k;p_R,P).
\end{equation}
Furthermore, $S[{}^1\!S_0^{(8)}](k;p_R,P)$ is normalized to the 
conventional NRQCD matrix element according to (\ref{soft1}), i.e. 
\begin{equation}
\label{norm1s}
\int\!\frac{d^3 \bff{k}}{(2\pi)^3 2 k^0} \,S[{}^1\!S_0^{(8)}](k;p_R,P) 
= \langle {\cal O}_8({}^1\!S_0)\rangle.
\end{equation}

\begin{figure}
  \vspace*{-0.2cm}
  \begin{center}
    \leavevmode
    \hspace*{1cm}
    \includegraphics[width=.4\textwidth,bb=135 100 540 390]{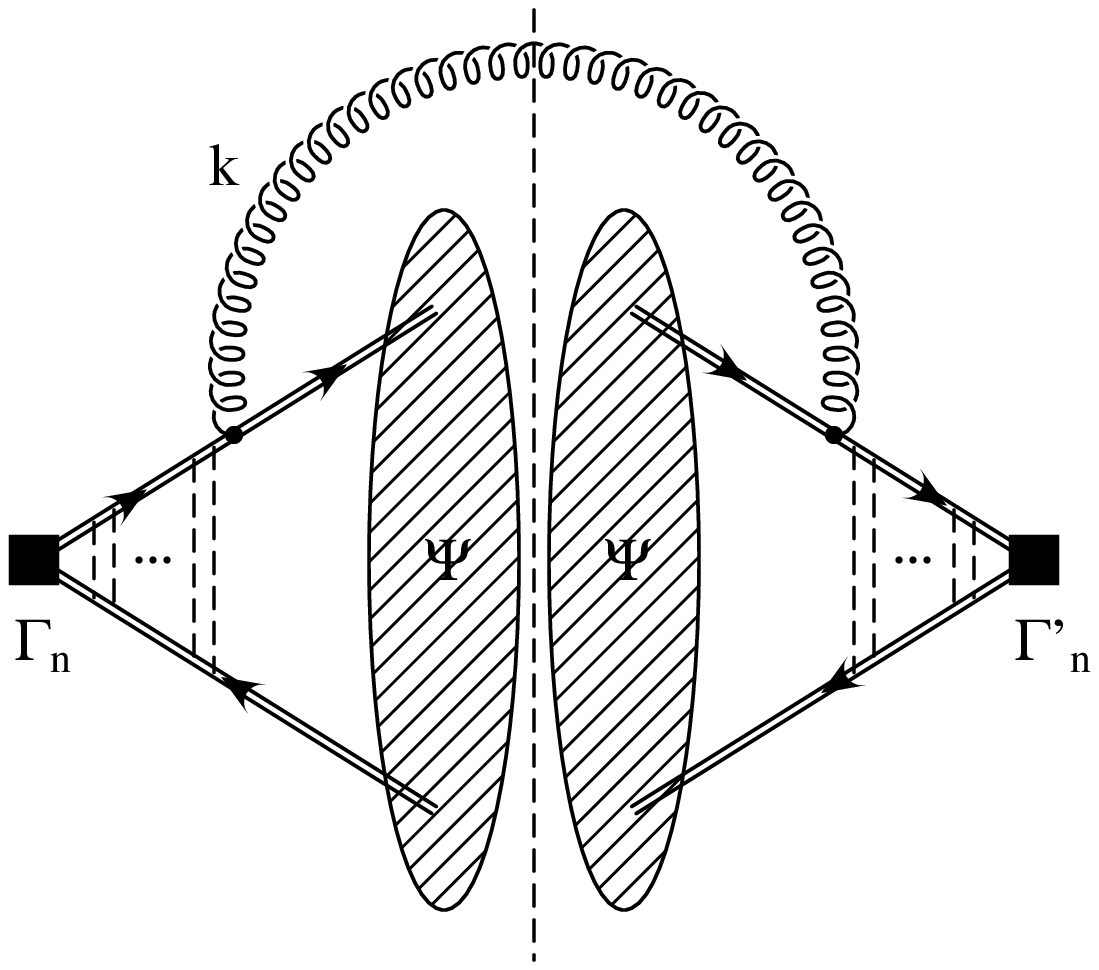}
    \hspace{.00\textwidth}
    \includegraphics[width=.4\textwidth,bb=135 100 540 390]{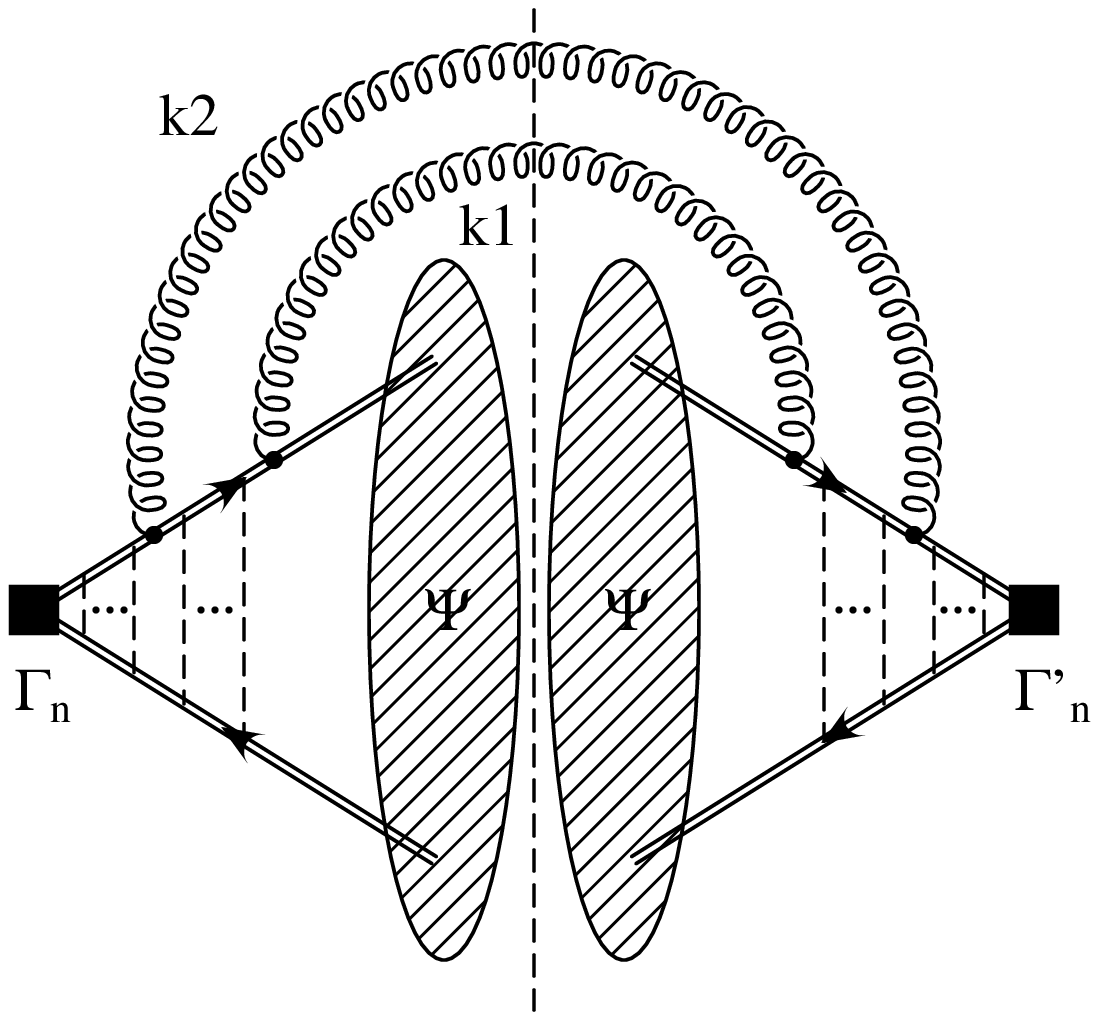}
  \end{center}
  \vspace*{0.2cm}
\caption{Left: one of four NRQCD diagrams which contribute to 
the M1 (E1) transition from 
an $^1\!S_0^{(8)}$ ($^3\!P_0^{(8)}$) state, specified by $\Gamma_n^{(')}$, 
to $J/\psi$. Right: an example for 
a double E1 transition from a  $^3\!S_1^{(1,\,8)}$ state.}
\label{nrqed}
\end{figure}
The non-relativistic quark-gluon vertices are classified according to 
their velocity suppression. The leading spin-flipping interaction is 
provided by the chromo-magnetic interaction vertex 
$-g_s/(2 m_c)(\bff{\sigma}\times \bff{k})$ with $k$ the outgoing 
gluon momentum as in Fig.~\ref{nrqed}. The diagram on the left hand side 
of Fig.~\ref{nrqed} gives (cf.~(\ref{ampl}))
\begin{eqnarray}
\label{firstdiag}
&& \frac{g_s^2 C_F}{8 m_c^2} 
\int\!\frac{d^3 \bff{k}}{(2\pi)^3 2 k^0} 
\left(\delta_{ij}-\frac{k_i k_j}{\bff{k}^2}\right)
\int\!\frac{d^3\bff{p}}{(2\pi)^3}\frac{d^3\bff{p}'}{(2\pi)^3}
\frac{d^3\bff{q}}{(2\pi)^3}\frac{d^3\bff{q}'}{(2\pi)^3}
\nonumber\\
&& \hspace*{1cm}
\cdot \,
\frac{1}{2}\sum_\lambda\mbox{tr}(\bff{\epsilon}(\lambda)\!\cdot\!\bff{\sigma} 
(\bff{\sigma}\times\bff{k})_i)\,
\mbox{tr}(\bff{\epsilon}^*(\lambda)\!\cdot\!\bff{\sigma} 
(\bff{\sigma}\times(-\bff{k}))_j)\,\psi(\bff{p})\,\psi(\bff{p'})
\nonumber\\
&& \hspace*{1cm}
\cdot \,i G_c(\bff{p}+\bff{k}/2,\bff{q}+\bff{k}/2;E(p_R+k))\,
i G_c(\bff{q}'+\bff{k}/2,\bff{p}'+\bff{k}/2;E(p_R+k)),
\end{eqnarray}
with $\psi(\bff{p})$ as given by (\ref{wavefn}) and 
$E(p_R+k)=p_R^0+k^0-2 m_c=-m_c (C_F\alpha_s)^2/4+k^0\equiv -\kappa^2/m_c$.
Eq.~(\ref{firstdiag}) can be simplified, because the gluon is 
ultrasoft with energy and momentum of order $m_c v^2\sim 
m_c \alpha_s^2$, while 
$\bff{p}$, $\bff{p}'$, $\bff{q}$ and $\bff{q}'$ are of 
order $m_c v\sim m_c \alpha_s$. Dropping small terms in the arguments of the 
Coulomb Green function (as we have already done when defining 
$E(p_R+k)$), performing the traces and accounting for an identical 
contribution from the other three diagrams not shown in Fig.~\ref{nrqed}, 
we obtain
\begin{equation}
\label{s1s}
S[{}^1\!S_0^{(8)}](k;p_R,P) = \frac{2 g_s^2 C_F}{m_c^2}\,
\bff{k}^2 \left| I[{}^1\!S_0^{(8)}](k) \right|^2,
\end{equation}
where
\begin{equation}
I[^1\!S_0^{(8)}](k)  = 
\int \!\frac{d^3\bff{q}}{(2\pi)^3} \frac{d^3\bff{p}}{(2\pi)^3}
\,G_c(\bff{q},\bff{p};-\kappa^2/m_c) \, \psi(\bff{p}).
\end{equation}

To compute this integral, we switch to coordinate space, 
\begin{equation}
\label{coord}
I[^1\!S_0^{(8)}](k) =
\int \! d^3\bff{x}\,\tilde{G}_c(\bff{x},\bff{0};-\kappa^2/m_c)\,
\tilde{\psi}(\bff{x}),
\end{equation}
use $\tilde{\psi}(\bff{x}) =
\sqrt{\gamma^3/\pi} e^{- \gamma x}$ ($\gamma=m_c C_F\alpha_s/2$), 
gained by Fourier 
transformation of (\ref{wavefn}), and the following representation for 
the coordinate space Coulomb Green function \cite{V79}\footnote{There 
is a misprint in the first reference of \cite{V79}, which is corrected 
in Eq.~(18) of the second reference.}:
\begin{equation}
\label{lgreen}
\tilde{G}_c(\bff{x}, \bff{y}; -\kappa^2/m_c)
= \sum_{l=0}^\infty \,(2l+1)\,
\tilde{G}_l(x, y; -\kappa^2/m_c)\,P_l(\bff{x}\!\cdot\!\bff{y}/(xy)),
\end{equation}
where $x$, $y$ denote the modulus of $\bff{x}$, $\bff{y}$, $P_l(z)$ the 
Legendre polynomials and 
\begin{equation}
\label{green} 
\tilde{G}_l(x, y; -\kappa^2/m_c)
= \frac{m_c \kappa}{2\pi}\,(2\kappa x)^l (2\kappa y)^l e^{- \kappa (x + y)}
\sum_{s=0}^\infty \frac{L_s^{(2l+1)}(2\kappa x) L_s^{(2l+1)}(2\kappa y) s!}
{(s + l + 1 - \lambda \gamma/\kappa) (s + 2l + 1)!}.
\end{equation}
Here $L_s^{(2l+1)}(z)$ refers to the Laguerre polynomials and the parameter 
$\lambda$ is defined such that the Green function corresponds to the Green 
function in the potential
\begin{equation}
V(r)=-\lambda \,\frac{C_F\alpha_s}{r}.
\end{equation}
Hence $\lambda=1$, if the intermediate $c\bar{c}$ pair propagates 
in a colour singlet state, and $\lambda=-1/(2 N_c C_F)=-1/8$, if it propagates 
in a colour-octet state, which is what we need here. Only the $l=0$ 
component of the Green function contributes to the integral (\ref{coord}). 
The remaining radial integration over Laguerre polynomials is easily 
executed as an integral over the generating function
\begin{equation}
e^{-z u/(1-u)}\,\frac{1}{(1-u)^{p+1}} = 
\sum_{s=0}^\infty u^s\,L_s^{p}(z)
\end{equation}
with subsequent expansion in $u$. Then, summing over $s$, 
and introducing the dimensionless variable 
\begin{equation}
z=\kappa/\gamma=\left(1+\frac{4k_0}{m_c C_F^2\alpha_s^2}\right)^{\!1/2},
\end{equation}
the result is 
\begin{eqnarray}
I[^1\!S_0^{(8)}](k) &=& 
-\frac{4m_c}{(\pi \gamma)^{1/2}} \frac{z^2}{(z^2 - 1)^2}
\sum_{s=1}^\infty \frac{s\,(s-1/z)}{s-\lambda/z}
\left( \frac{1-z}{1+z} \right)^s
\nonumber\\
&&\hspace*{-1.6cm}
=\,\frac{m_c}{(\pi\gamma)^{1/2}}\,\frac{1}{z^2-1}\,
\Bigg\{\,1+(\lambda-1)\bigg[\frac{2}{z+1}
\nonumber\\
&&\hspace*{-1cm}
-\,\frac{4 z}{z^2-1}\Big(1-{}_2F_1(-\lambda/z,1,1-\lambda/z;
(1-z)/(1+z))\Big)\bigg]\Bigg\}
\end{eqnarray}
with ${}_2F_1(a,b,c;z)$ the hypergeometric function. Let us check the 
power counting: with $\gamma\sim m_c v$ and $k_0, \,k_i \sim m_c v^2$, 
we obtain 
$I[^1\!S_0^{(8)}](k)\sim (m_c/v)^{1/2}$ and, from (\ref{norm1s}), (\ref{s1s}), 
$\langle {\cal O}_8({}^1\!S_0)\rangle\sim \alpha_s m_c^3 v^7$. This agrees 
with the velocity power counting of \cite{BBL95}. The additional $\alpha_s$ 
arises, because we consider the weak coupling limit.

The chromo-electric dipole transition 
$c\bar{c}[{}^3\!P_0^{(8)}]\to J/\psi+g$ is computed along similar lines. 
We have 
\begin{equation}
\label{s3p}
S[{}^3\!P_0^{(8)}](k;p_R,P) = \frac{8 g_s^2 C_F}{3 m_c^4}\,
\left| I[{}^3\!P_0^{(8)}](k) \right|^2,
\end{equation}
where
\begin{equation}
\label{coord3p}
I[^3\!P_0^{(8)}](k) =
\frac{1}{3}\int \! d^3\bff{x}\,\left[\frac{\partial}{\partial y_i}\,
\tilde{G}_c(\bff{x},\bff{y};-\kappa^2/m_c)\right]_{\bff{y}=0}\,
\frac{\partial}{\partial x_i}\tilde{\psi}(\bff{x}),
\end{equation}
and the normalization is given by
\begin{equation}
\label{norm3p}
\int\!\frac{d^3 \bff{k}}{(2\pi)^3 2 k^0} \,S[{}^3\!P_0^{(8)}](k;p_R,P) 
= \frac{\langle {\cal O}_8({}^3\!P_0)\rangle}{m_c^2}.
\end{equation}
The derivatives in (\ref{coord3p}) come from the factor $\bff{p}$ 
in the electric dipole vertex $-ig_s(\bff{p}+\bff{p}')/(2 m_c)\approx 
(-i)g_s \bff{p}/m_c$. 
In this case only the $l=1$ component of the Green function survives 
the $\bff{y}\to 0$ limit and the angular integration. The result is 
\begin{eqnarray}
I[^3\!P_0^{(8)}](k) &=& 
-\frac{4 m_c\gamma^{3/2}}{3\pi^{1/2}}\,\frac{z^3}{(z+1)^4}
\sum_{s=0}^\infty\frac{(s+1)(s+2)(s+3)}{s+2-\lambda/z}\,\left(\frac{1-z}
{1+z}\right)^{\!s}
\nonumber\\
&&\hspace*{-1.5cm}=\,\,
-\frac{m_c\gamma^{3/2}}{3\pi^{1/2}}\,\frac{1}{(z+1)^3}
\,\Bigg\{2 (1+z)(2+z)+(\lambda-1) (5+3 z)+2 (\lambda-1)^2
\nonumber\\
&&\hspace*{-0.8cm}
+\,\frac{4 z (1+z)(z^2-\lambda^2)}{(1-z)^2} \bigg[
{}_2F_1(-\lambda/z,1,1-\lambda/z;(1-z)/(1+z))-1
\nonumber\\
&&\hspace*{-0.8cm}+\,\frac{\lambda (1-z)}
{(1+z)(z-\lambda)}\bigg]\Bigg\}.
\end{eqnarray}
Velocity power counting gives $\langle {\cal O}_8({}^3\!P_0)\rangle/m_c^2 
\sim \alpha_s m_c^3 v^7$, which is again consistent with the standard 
counting.

\begin{figure}[t]
   \vspace{-3.7cm}
   \epsfysize=13.5cm
   \epsfxsize=9cm
   \centerline{\epsffile{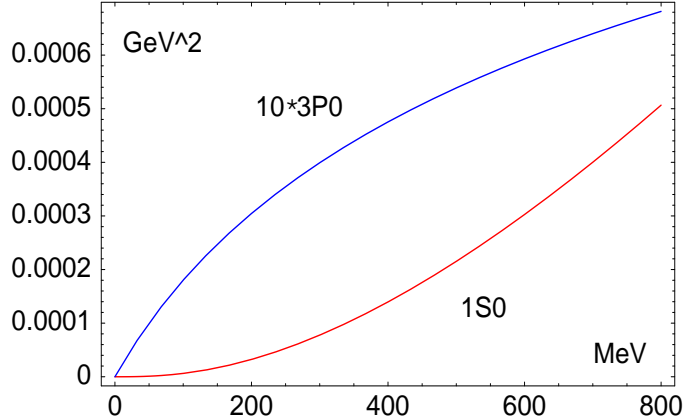}}
   \vspace*{-3.3cm}
\caption[dummy]{\label{fig4}\small 
Dependence of $k_0/(4\pi^2)\,S_n(k;p_R,P)$ for $n={}^1\!S_0^{(8)}$ and 
$n={}^3\!P_0^{(8)}$ on the emitted gluon energy $k_0$. The parameters 
are chosen as $m_c=1.5\,$GeV, $\alpha_s=0.4$, $\lambda=-1/8$.}
\end{figure}
The dependence of 
$k_0/(4\pi^2)\,S_n(k;p_R,P)$ for $n={}^1\!S_0^{(8)}$ and 
$n={}^3\!P_0^{(8)}$ on the energy $k_0$ of the emitted gluon is 
shown in Fig.~\ref{fig4}. 
The input parameters are chosen as $m_c=1.5\,$GeV, $\alpha_s=0.4$;  
$\lambda=-1/8$ for a colour octet matrix element. Both dependences 
are smooth and mainly reflect the asymptotic behaviours at small 
and large gluon energy. In particular the suppression of the 
${}^1\!S_0$ curve at small $k_0$ is a consequence of the structure of 
the magnetic dipole vertex. 

According to the normalization conditions (\ref{norm1s}) and (\ref{norm3p}) 
the integration of the two curves gives the value of the conventional 
NRQCD matrix elements. The result depends strongly (see the discussion 
below) on the cut-off on the integration range for $k_0$. Choosing the 
cut-off between $300\,$MeV and $600\,$MeV, we find\footnote{These numbers, 
in particular the one for ${}^3\!P_0$, depend sensitively on $\alpha_s$. 
$\langle {\cal O}_8^{J/\psi}({}^3\!P_0)\rangle/m_c^2$ increases 
rapidly as $\alpha_s$ increases.}
\begin{eqnarray}
\langle {\cal O}_8^{J/\psi}({}^1\!S_0)\rangle &=& (0.07-0.61)\cdot 10^{-4}\,
\mbox{GeV}^3
\nonumber\\
\langle {\cal O}_8^{J/\psi}({}^3\!P_0)\rangle/m_c^2 &=& 
(0.07-0.22)\cdot 10^{-4}\,
\mbox{GeV}^3.
\end{eqnarray}
Although these numbers may be insignificant, because the assumption 
$m_c v^2\gg \Lambda_{\rm QCD}$ necessary to obtain them, is not valid
for charmonium, it is interesting to note that the matrix elements come 
out one to two orders of magnitude smaller than the 
phenomenological values, determined from fitting colour-octet subprocesses 
to experimental data \cite{revs}. This suggests either a large 
non-perturbative enhancement of the matrix elements -- such as the presence 
of a gluon condensate to which the soft gluons can couple -- 
or the possibility 
that the phenomenological values of the matrix elements effectively 
parametrize other corrections to the production processes not related 
to soft gluon emission (such as higher order short-distance corrections).

The behaviour of the soft function $S_n(k;p_R,P)$ at large $k_0$ 
deserves further discussion. First, we observe that the calculation by 
itself does not provide an intrinsic cut-off for large $k_0$. This 
should not be expected, since at the level of perturbative radiation 
the ultraviolet behaviour of the soft function joins smoothly to the 
infrared behaviour of the short-distance part. A well known example of 
this occurs in $P$-wave production \cite{BBL95}: 
the logarithmic infrared behaviour 
of the coefficient function of $\langle {\cal O}_1^\chi(^3\!P_0)\rangle$ 
matches the logarithmic ultraviolet divergence of 
$\langle {\cal O}_8^\chi(^3\!S_1)\rangle$. 

Inspection of 
$S[^1\!S_0^{(8)}](k;p_R,P)$ shows that we obtain a quadratically 
ultraviolet divergent matrix element 
$\langle {\cal O}_8^{J/\psi}(^1\!S_0)\rangle$, which seems to contradict 
the conventional wisdom that this matrix element is scale-independent 
at leading order. However, the conventional wisdom is derived from the 
use of dimensional regularization. If a hard cut-off on the gluon energy 
is used, the colour octet $^1\!S_0$ operator mixes into the colour 
singlet $^3\!S_1$ operator through a quadratically divergent term.\footnote{
The dimensions 
work out correctly, because the two chromo-magnetic 
dipole vertices provide two powers of $1/m_c$.} This corresponds to a 
infrared finite, but quadratically infrared sensitive contribution to 
the coefficient function of 
$\langle {\cal O}_1^{J/\psi}(^3\!S_1)\rangle$, consistent with the 
over-all $v^4$ suppression of $\langle {\cal O}_8^{J/\psi}(^1\!S_0)\rangle$ 
relative to $\langle {\cal O}_1^{J/\psi}(^3\!S_1)\rangle$. In dimensional 
regularization, the quadratically infrared sensitive term is attributed 
entirely to the short-distance coefficient and the quadratic divergence 
in $\langle {\cal O}_8^{J/\psi}(^1\!S_0)\rangle$ is set to zero.

In case of $S[^3\!P_0^{(8)}](k;p_R,P)$ we find a linear divergence for 
$\langle {\cal O}_8^{J/\psi}(^3\!P_0)\rangle$. The interpretation of this 
divergence requires a more careful discussion of the $k_0$-integral and 
the integrals over $\bff{p}$ and $\bff{p}'$; it will not be presented 
here.

In the ansatz (\ref{ansatz}) we have added a cut-off on $k_0$ by hand 
in the form of an exponential fall-off for $k_0\gg\Lambda_n$. We 
interpret this ansatz as a `primordial distribution' for the radiation of 
non-perturbative gluons, which eventually is modified by perturbative 
evolution. This is similar to the assumption that intrinsic 
transverse momenta of the proton's constituents are bounded. Perturbative 
radiation violates this assumption and leads to the evolution of 
parton distributions. A similar ansatz is also implied by the ACCMM model 
or in shape functions for semileptonic $B$ decays in general.

Finally, we comment on the transition from a colour octet or a colour 
singlet $^3\!S_1$ state to $J/\psi$. 
This presents a more complicated case, since -- besides the contribution 
with no gluon emission for the colour singlet state -- the leading term 
requires the emission of two gluons, see the right hand side of 
Fig.~\ref{nrqed}. In coordinate space this requires the evaluation of 
integrals of the form  
\begin{eqnarray}
\label{IntE1E1}
I_{ij}[^3\!S_1^{(1,\,8)}] &\sim& 
\int \! d^3\bff{x} d^3\bff{y}\,\tilde{G}_c(-\bff{y},\bff{0};E(p_R+k_1))
\nonumber\\
&&\hspace{2em} 
\times\left( \frac{\partial}{\partial y_i}
\tilde{G}_c(\bff{x},\bff{y};E(p_R+k_1+k_2))\right)
\left( \frac{\partial}{\partial x_j} \tilde{\psi}(\bff{x})\right),
\end{eqnarray}
which we shall not pursue. 
If we were only interested in the limit of small loop
momenta $k = k_1 + k_2$, we could expand the Green functions for $k^0_i
\ll \gamma^2/m_c$ first and integrate afterwards over $\bff{x}$ and
$\bff{y}$. We would then find the same small-$k_0$ behaviour as 
in the case of $I[^3\!P_0](k)$. 

\section{Momentum spectrum in $B\to J/\psi X$}
\label{bsect}

\noindent
In this section we apply the formalism developed 
in the previous section to 
the $J/\psi$ momentum spectrum in the semi-inclusive decay 
$B \to J/\psi X$.  The leading partonic decay process is very simple, 
resulting in $J/\psi$ with fixed momentum, but the hadronic decay 
spectrum is modified by  fragmentation of the $c\bar{c}$ pair, which 
is the main concern of this paper, and by bound state effects on the 
$b$ quark in the $B$ meson. Both will be taken into account in the 
following. 

We start by recapitulating the partonic result for $b \to c\bar{c}[n]
+ q$. We then implement the fragmentation of the $c\bar{c}$ pair 
according to our shape function ansatz and obtain the $J/\psi$ 
momentum distribution in $b$ quark decay. We regard this distribution 
as input distribution for the ACCMM model, which accounts in a 
simple but satisfactory way for the effect of Fermi motion of the $b$ 
quark inside the $B$ meson. The resulting $J/\psi$ distribution in 
$B$ meson decay is then boosted to the CLEO frame and compared 
to CLEO data. The aim of this comparison is twofold: first we show 
that smearing of the spectrum due to fragmentation of the $c\bar{c}$ 
pair is essential to describe the CLEO data. Second we use these data 
to determine the shape function model parameter $\Lambda$. Assuming 
universality of the shape function over the whole kinematic domain, 
we will then turn to $J/\psi$ photoproduction in 
Sect.~\ref{photoproduction}. Results for the $J/\psi$ momentum 
distributions already exist in the literature, including colour octet 
production \cite{PPS97,AH99}. However, only Fermi motion effects are taken 
into account there. We will briefly compare our results with 
those of  \cite{PPS97,AH99} at the end of this section.

\subsection{Energy distribution in $b$ quark decay}

\noindent
The underlying partonic process of a $B$ meson 
decay into $J/\psi$ and light
hadrons is $b \to c\bar{c}[n] + q$ ($q = \{d, s\}$). Since the $c\bar{c}$ pair
is treated as a single particle kinematically a leading order calculation of
this process results in a fixed value for its energy (momentum) rather than
in a real spectrum. Defining\footnote{In this section ``hatted'' quantities 
refer to the $b$ quark rest frame.}  $\hat{z}=2 \hat{E}_{c\bar{c}}/m_b$ as the 
energy fraction 
of the $c\bar{c}$ pair in the $b$ quark rest frame, the ``spectrum'' is 
\begin{equation}
\frac{d\Gamma_{\!\!c\bar{c}}}{d\hat{z}}
= \Gamma_{\!\!c\bar{c}} \, \delta(1 + \eta - \hat{z}) \,,
\end{equation}
where $\eta = 4m_c^2/m_b^2$ for massless light hadrons in the 
final state. In a purely partonic calculation one may identify $2 m_c$ 
with the $J/\psi$ mass and $m_b$ with the $B$ meson mass.

At leading order in the non-relativistic expansion the $c\bar{c}$ pair 
has to be produced in a colour singlet ${}^3 S_1$ state. This term 
coincides with the colour singlet model and has been computed long 
ago  \cite{DT80,Wise80}. At relative order $v^4\approx 1/15$ in 
the non-relativistic expansion, $J/\psi$ can also be produced through 
$c\bar{c}$ in ${}^1S_0^{(8)}, \, ^3P_J^{(8)}, \, ^3S_1^{(8)} $ colour 
octet states. These formally subleading contributions are enhanced 
by a factor of about $15$, by which the short-distance structure of the 
$\Delta B=1$ weak effective Hamiltonian favours the production of 
colour octet $c\bar{c}$ pairs in the $b\to c\bar{c} q$ transition. These 
additional terms can be comparable or even larger than the colour 
singlet term \cite{KLS96a,KLS96b,BMR99}. They are the ones of interest in this 
paper, since the radiation of soft gluons in colour octet $c\bar{c}$ 
fragmentation has a large kinematic effect on the observed $J/\psi$ 
momentum spectrum. In comparison, fragmentation effects in the 
colour singlet channel are order $v^4$ suppressed relative to 
the total colour singlet rate and therefore negligible. Hard perturbative 
corrections to the colour singlet \cite{BMR99,BE94} and colour octet 
\cite{BMR99} production processes are also known. They enhance 
the colour octet channels moderately. Within the present limitations of 
the shape function ansatz we must neglect these perturbative corrections 
for consistency.
The partonic production spectra for the $c\bar{c}[n]$ states of 
interest read 
\begin{equation}
\frac{d\Gamma_{\!\!c\bar{c}}[n]}{d\hat{z}} =
\frac{1}{2m_b} \frac{1-\eta}{8\pi} \,H_n(m_b, 2m_c) \,
\delta(1 + \eta - 
\hat{z}),
\end{equation}
where 
\begin{equation}
\label{hardB}
H_n(m_b, 2m_c)
= \frac{2 G_F^2 |V_{cb}|^2 m_b^4}{27\pi (2m_c)} \,C_{[1,8]}^2 f[n](\eta)
\end{equation}
and the process-specific functions $f[n](\eta)$ are given by 
\cite{KLS96a,KLS96b,BMR99} 
\begin{eqnarray}
f[^3S_1^{(1)}](\eta) &=& (1 - \eta) (1 + 2\eta) ,
\\
f[^3S_1^{(8)}](\eta) &=& \frac{3}{2} (1 - \eta) (1 + 2\eta) ,
\\
f[^1S_0^{(8)}](\eta) &=& \frac{9}{2} (1 - \eta) ,
\\
f[^3P_J^{(8)}](\eta) &=& 9 (1 - \eta) (1 + 2\eta) .
\end{eqnarray}
Note that the colour octet matrix elements are not part of the hard 
amplitudes, but included in the normalization of the radiation 
function $\Phi_n(k)$, see (\ref{normalization}). In case of the 
$P$ wave contribution, the normalization refers to $\langle {\cal O}_8
({}^3P_0)\rangle/m_c^2$ and the corresponding factor $1/m_c^2$ is 
also extracted from $H[^3P_J^{(8)}](m_b,2m_c)$. 
As mentioned above we neglect QCD corrections and also small 
corrections due to penguin operators. The Wilson coefficients 
$C_{[1,8]}$ of the effective operators in the
weak $\Delta B = 1$ Hamiltonian are related to the usual $C_\pm$ by
\begin{eqnarray}
C_{[1]}(\mu) &=& 2C_+(\mu) - C_-(\mu),
\nonumber\\
C_{[8]}(\mu) &=& C_+(\mu) + C_-(\mu).
\end{eqnarray}
At leading order, as appropriate to the present analysis,
\begin{equation}
C_\pm(\mu) = \left[ \frac{\alpha_s(M_W)}{\alpha_s(\mu)}
                                        \right]^{\gamma_\pm^{(0)}/(2\beta_0)}
\end{equation}
with
$\gamma_\pm^{(0)} = \mbox{}\pm 2 (3 \mp 1) $ and $
\beta_0 = 11 - 2n_f/3$. In (\ref{hardB}) the notation $C_{[1,8]}$ implies 
$C_{[1]}$, if $n$ is a colour-singlet state, and $C_{[8]}$, if $n$ is a colour 
octet state. Note that $C_{[8]}^2/C_{[1]}^2\approx 15$ at $\mu\sim m_b$.

We now implement $c\bar{c}$ fragmentation for the colour octet 
production channels. Notice that the partonic amplitude
squared has no azimuthal dependence, hence 
$\bar{H}_n(m_b, 2m_c) = H_n(m_b, 2m_c)$ in the notation of (\ref{azav}). 
Furthermore, we need the light cone components of the
incoming momentum $\hat{P}_{in} = (m_b, \bff{0})$ in the $J/\psi$ rest frame
to get $\alpha$ and $\beta$ of (\ref{albe}). We find 
\begin{equation}
\label{albeB}
\alpha = \frac{m_b}{M_R} (\hat{E}_R - |\hat{\bff{p}}_R|) - M_R,
\qquad
\beta = \frac{m_b}{M_R} (\hat{E}_R + |\hat{\bff{p}}_R|) - M_R.
\end{equation}
The index ``R'' now refers to $J/\psi$. To complete the implementation we 
have to fix the ambiguity in treating the kinematic effects in the hard 
production amplitudes $H_n$. Strictly speaking the shape function 
formalism allows us to ignore the dependence of the hard production 
process on the vector $l$, since it does not lead to singular contributions 
near the endpoint, if the hard matrix element is not singular at the 
endpoint. On the other hand, the invariant mass of the $c\bar{c}$ pair 
is kinematically given by 
\begin{equation}
M^2_{c\bar{c}} (k) = (p+l)^2 = (p_R+k)^2 = M_R^2+2 M_R k_0+k^2,
\end{equation}
where $k$ is the four momentum of soft radiation in the $J/\psi$ rest 
frame. We adopt the convention that $2 m_c$ in the partonic 
matrix element is replaced by $M_{c\bar{c}}(k)$ everywhere, i.e. 
even when it does not arise kinematically, but through internal charm 
quark propagators. This convention is consistent with the shape function 
formalism in the shape function limit, but is arbitrary otherwise. It has 
the advantage of incorporating the physically expected effect of 
reducing the short-distance amplitude, because of the need to 
create a heavier $c\bar{c}$ pair as compared to a purely partonic 
picture. The only exception to the convention is the factor 
$1/(2 m_c)$ in (\ref{hardB}), which comes from the normalization of 
the $c\bar{c}$ state. It should be replaced by $1/M_R$.  
Eq.~(\ref{master}), specialized to the $J/\psi$ energy distribution in 
$b$ quark decay, is then:
\begin{equation}
\label{bspectrum}
\frac{d\hat{\Gamma}}{d\hat{E}_R} = \frac{|\hat{\bff{p}}_R|}{4\pi^2}
\sum_n \int\limits_0^{\alpha\beta} \frac{dk^2}{2\pi}\!
\int\limits_{(\alpha^2+k^2)/(2\alpha)}^{(\beta^2+k^2)/(2\beta)}
\!\!\!dk_0\,\,
\frac{1}{2 m_b} \,H_n(m_b,M_{c\bar{c}}(k))\,
\frac{M_R}{8\pi m_b |\hat{\bff{p}}_R|}\,\Phi_n(k) 
\end{equation}
with $\alpha$, $\beta$ from (\ref{albeB}).

\subsection{Normalization difficulty}

\noindent
We assumed up to now that the radiation function $\Phi_n(k)$ is 
normalized according to (\ref{normalization}). This implies that as 
$\Lambda$ of (\ref{ansatz}) tends to zero the integral  
over $d\hat{\Gamma}/d\hat{E}_R$ of (\ref{bspectrum}) 
equals the integrated partonic rate with $m_c=M_R/2$. 

Consider now the integral $\hat{\Gamma}_n(\Lambda)$ of the 
spectrum (\ref{bspectrum}) with fragmentation (for a specific production 
channel $n$) at small $\Lambda$ and expand in $\Lambda$. To make things 
simpler put $k=0$ in the hard matrix element $H_n$. Then integrate over 
$\hat{E}_R$ or, equivalently, $\bar{z}\equiv 2\hat{E_R}/m_b$, and 
perform a change of variables 
from $\bar{z}$ to $\alpha$. Then note that for small $\Lambda$, one 
can set the upper limits of the $k^2$ and $k_0$ integrations to infinity up 
to exponentially small corrections in $\Lambda$, given the ansatz 
(\ref{ansatz}). Then exchange the $k^2$ and $k_0$ integration 
with the $\alpha$ integration to obtain
\begin{equation}
\label{gam1}
\hat{\Gamma}_n(\Lambda) = \frac{M_R}{16 (2\pi)^4 m_b} H_n(m_b,M_R)\int
\limits_0^\infty dk^2\int\limits_{\sqrt{k^2}}^\infty 
d k_0 \,\Phi_n(k)\!\!\!
\int\limits_{k_0-\sqrt{k_0^2-k^2}}^{k_0+\sqrt{k_0^2-k^2}} \!\!\!
d\alpha\,\left|\frac{d\bar{z}}{d\alpha}\right|.
\end{equation}
Now introduce the average
\begin{equation}
\label{nav}
\langle\langle f\rangle\rangle_n \equiv \frac{1}{(2\pi)^3}
\,\frac{1}{\langle {\cal O}_n^{J/\psi}\rangle}\,\int
\limits_0^\infty dk^2\int\limits_{\sqrt{k^2}}^\infty 
d k_0 \,\sqrt{k_0^2-k^2}\,\Phi_n(k)\,f(k),
\end{equation}
defined such that $\langle\langle 1\rangle\rangle_n =1$ according to the 
normalization condition (\ref{normalization}). Eq.~(\ref{gam1}) is 
then rewritten in the form 
\begin{equation}
\hat{\Gamma}_n(\Lambda) = \frac{1}{2m_b} \frac{1-\eta}{8\pi} \,
H_n(m_b, M_R) \, \langle {\cal O}_n^{J/\psi}\rangle \cdot 
r_n(\Lambda),
\end{equation}
where $\eta$ is now defined as $M_R^2/m_b^2$ and 
\begin{equation}
r_n(\Lambda) = \frac{M_R}{2(1-\eta)}\,\langle\langle \,\frac{1}
{\sqrt{k_0^2-k^2}}
\!\!\!\!
\int\limits_{k_0-\sqrt{k_0^2-k^2}}^{k_0+\sqrt{k_0^2-k^2}} \!\!\!\!\!\!
d\alpha\,\left|\frac{d\bar{z}}{d\alpha}\right|\,\rangle\rangle_n.
\end{equation}
Hence we obtain the partonic decay rate with $m_c=M_R/2$ up to the 
factor $r_n(\Lambda)$. To evaluate $r_n(\Lambda)$ in an expansion 
in $\Lambda$ we observe that 
\begin{equation}
\left|\frac{d\bar{z}}{d\alpha}\right| = \frac{1}{M_R}\left(
\frac{1}{(1+\alpha/M_R)^2}-\eta\right) = \frac{1}{M_R}
\left(1-\eta+\sum_{n=1}^\infty (n+1)\left(-\frac{\alpha}{M_R}\right)^n\right)
\end{equation}
can be expanded under the integral. The result is
\begin{equation}
r_n(\Lambda) = 1 + \frac{1}{1-\eta}\left(-2\,\langle\langle \,\frac{k_0}{M_R}
\,\rangle\rangle_n+\langle\langle \,\frac{4 k_0^2-k^2}
{M_R^2}\,\rangle\rangle_n +O\left(\frac{\Lambda^3}{M_R^3}\right)\right).
\end{equation}
The averages can be done using (\ref{ansatz}) with $a_n$ fixed by 
(\ref{normalization}); they scale with definite powers of $\Lambda$ as 
follows from the form of (\ref{nav}). With $\eta=0.416$, the result is
\begin{equation}
\label{rn1}
r_n(\Lambda) = 1 - \left\{\begin{array}{cc}4.76\\5.75\end{array}\right\}
\frac{\Lambda}{M_R} + \left\{\begin{array}{cc} 12.97\\19.53\end{array}\right\}
\left(\frac{\Lambda}{M_R}\right)^2 + \ldots,
\end{equation}
where the upper number refers to $b_n=0$ in (\ref{ansatz}) and the lower 
one to $b_n=2$. 
For $\Lambda\approx 300\,$MeV this implies large corrections to the 
integrated rate. Since $\Lambda \sim m_c v^2$, this must be interpreted 
as large higher order corrections in the velocity expansion, which are 
not taken into account in the usual leading order NRQCD analysis. This 
means that enforcing the normalization condition (\ref{normalization}) 
underestimates the data, because the matrix elements on the right hand 
side of (\ref{normalization}) have been obtained without these large 
higher order corrections. 

The effect is in fact even larger than indicated by (\ref{rn1}), 
because we keep the $k$ dependence of the hard matrix element and 
$M_{c\bar{c}}(k)$ is always larger than $M_R$. As an indication of this 
effect we can compute the average
\begin{equation}
\label{meff}
4 {m_c^{\rm eff}}^2\equiv \langle\langle \,
M_{c\bar{c}}(k)^2 \,\rangle\rangle_n \approx 
M_R^2\left(1+\left\{\begin{array}{cc}2.78\\3.39\end{array}\right\}
\frac{\Lambda}{M_R}\right)
\end{equation}
which implies an effective charm quark mass of about $1.8\,$GeV 
rather than $m_c=1.5\,$GeV which is usually adopted in partonic 
NRQCD calculations. 

When the implicit $k$-dependence of the 
partonic matrix element $H_n$ is taken into account, the numbers given 
in (\ref{rn1}) change. However, the observation that 
$v^2$ corrections are large is generic.

\subsection{Fermi motion effects}

\noindent 
We now convert the spectrum (\ref{bspectrum}) 
in $b$ quark decay into a spectrum in 
$B$ meson decay by accounting for Fermi motion of the $b$ quark. 
We make the reasonable assumption that $B$ meson bound state effects 
can be factorized from the hard subprocess as well as from $c\bar{c}$ 
fragmentation. The Fermi motion effect can be described rigorously 
in heavy quark effective theory \cite{N94}, but we contend ourselves 
with the earlier ACCMM model \cite{ACM}. The ACCMM model is in 
fact consistent with the heavy quark expansion, if a particular relation 
between the $b$ quark mass and the ACCMM model parameter 
$p_F$ is adopted \cite{CR94}. (The ACCMM model then 
assumes a particular 
value for the kinetic energy matrix element of heavy quark effective 
theory.) The ACCMM model provides a phenomenologically viable 
description of energy spectra in other $B$ decays, 
e.g.~$B \to X \ell \bar{\nu}_\ell$ or $B \to X_s\gamma$.

The basic idea of this model is quite intuitive: one imagines the $b$ quark
moving inside the $B$ meson at rest with a momentum $p$ according to 
some distribution with a width of a few hundred MeV. The cloud of gluons 
and light quarks in the $B$ meson of the mass $M_B$ is treated as 
spectator quark with mass $m_{sp}$. To keep the kinematics of this 
``decay in flight'' exact one introduces a so-called floating $b$ quark mass 
\begin{equation}
m_b^2(p) = M_B^2 + m_{sp}^2 - 2 M_B \sqrt{m_{sp}^2 + p^2}.
\end{equation}
The $b$ quark is on-shell with energy $E_b(p)=(m_b^2(p)+p^2)^{1/2}$. 
The $b$ quark momentum distribution must be chosen ad hoc. Usually 
one takes a properly normalized Gaussian form
\begin{equation}
\Phi_{\rm ACM}(p) = \frac{4}{\sqrt{\pi} p_F^3} \exp\,(- p^2/p_F^2),
\end{equation}
where $\int_0^\infty dp \,p^2\, \Phi_{\rm ACM}(p) = 1$. Implementing the 
kinematics of decay in flight, the $J/\psi$ energy distribution in the 
$B$ meson rest frame (quantities without ``hat'') is then obtained from the 
spectrum in $b$ quark decay (\ref{bspectrum}) by the convolution 
\begin{equation}
\label{BtoPsiX}
\frac{d\Gamma}{dE_R} =
\int\limits_{\max\{0, \, p_-\}}^{p_+} \!\!\!\!\!dp \, p^2\,
\Phi_{\rm ACM}(p) \frac{m_b^2(p)}{2 p E_b(p)}
\int\limits_{\hat{E}_R^{\rm min}(p)}^{\hat{E}_R^{\rm max}(p)}
\frac{d\hat{E}_R}{\hat{E}_R} \frac{d\hat{\Gamma}}{d\hat{E}_R} .
\end{equation}
The integration over the $J/\psi$ energy $\hat{E}_R$ in the $b$
quark rest frame is limited by
\begin{eqnarray}
\hat{E}_R^{\rm max} &=& 
\min\left\{ \frac{E_R E_b(p) + |\bff{p}_R| p}{m_b(p)},
        \, \frac{m_b^2(p)+M_R^2}{2m_b(p)} \right\} ,
\\
\hat{E}_R^{\rm min} &=& \frac{E_R E_b(p) - |\bff{p}_R| p}
{m_b(p)}. 
\end{eqnarray}
The requirement $\hat{E}_R^{\rm min} \leq {E}_R^{\rm max}$ leads to 
the following bounds on $p$:
\begin{equation}
p_\pm = \frac{[p_R \pm (M_B - E_R)]^2 - m_{sp}^2}
                   {2 [p_R \pm (M_B - E_R)]} .
\end{equation}

The dependence of the energy spectrum  (\ref{BtoPsiX}) on the 
two parameters of the ACCMM model, $m_{sp}$ and $p_F$, is quite 
different. Changing the
value of the spectator mass does not affect the spectrum noticeably. 
Therefore $m_{sp}$ usually is fixed to 150 MeV in all ACCMM analyses. 
On the other hand
the width $p_F$ of the momentum distribution must be chosen carefully,
because the shape of the spectrum is strongly sensitive to this parameter.
Successful fits to the lepton energy spectrum in semi-leptonic decay 
typically find $p_F \approx (300-450)\,$ MeV \cite{CLEOpF}.

\subsection{Final result and comparison with CLEO data}
\label{finalresult}

\noindent 
Eq.~(\ref{BtoPsiX}) yields the $J/\psi$ energy spectrum for $B$ mesons 
decaying at rest. To compare with CLEO data \cite{CLEO95}, we have 
to translate the energy spectrum (\ref{BtoPsiX}) into a momentum
spectrum
\begin{equation}
\frac{d\Gamma}{dp_R} = \frac{E_R}{p_R} \, \frac{d\Gamma}{dE_R}
\end{equation}
and account for the fact that $B$ mesons have momentum 
$\tilde{p}_B = (M_{\Upsilon(4 S)}^2/4 - M_B^2)^{1/2} \approx 482\,$MeV in the 
CLEO rest frame in which the data in \cite{CLEO95} is 
presented.\footnote{Quantities with ``tildes'' refer to the CLEO or 
$\Upsilon(4S)$ rest frame.}  The final boost from the $B$ meson to 
the $\Upsilon(4S)$ rest frame is effected by 
\begin{equation}
\frac{d\tilde{\Gamma}}{d\tilde{p}_R}
= \frac{\tilde{p}_R}{\tilde{E}_R} \, \frac{M_B}{2\tilde{p}_B}
\int\limits_{p_R^{\rm min}}^{p_R^{\rm max}} \!\!
\frac{dp_R}{p_R} \, \frac{d\Gamma}{dp_R} \,,
\end{equation}
where the bounds on the $J/\psi$ momentum
\begin{eqnarray}
p_R^{\rm min} &=& \max\left\{
0, \frac{\tilde{E}_B \tilde{p}_R - \tilde{p}_B \tilde{E}_R}{M_B} \right\}
\,, \\
p_R^{\rm max} &=& \min\left\{
\frac{\lambda^{1/2}(M_B^2, M_R^2, m_{sp}^2)}{2 M_B},
\frac{\tilde{E}_B \tilde{p}_R + \tilde{p}_B \tilde{E}_R}{M_B} \right\}
\end{eqnarray}
stem from kinematical restrictions set by the masses in the K\"allen function
$\lambda(x,y,z) = x^2 + y^2 + z^2 - 2xy - 2xz - 2yz$ and from the integration
over the angle between the $B$ and the $J/\psi$ momentum.

Owing to the difficulties of normalizing the partial production rates 
discussed above we forsake the idea of predicting the absolute  
$J/\psi$ branching fraction in $B$ decay and concentrate on the 
shape of the spectrum. We fix the absolute normalization by adjusting 
the sum of all contributions to data. This is actually equivalent to 
re-fitting the NRQCD matrix elements to data after including large 
higher order corrections in the velocity expansion. However, we do not 
give the result of the re-fitting, because we believe it is of little interest 
for comparison with other $J/\psi$ production processes.

The shape function ansatz (\ref{ansatz}) is slightly different for the
different production channels because of the different choice of 
parameters (\ref{bs}), (\ref{cutparam}). Therefore the shape of the 
momentum spectrum depends somewhat on the relative contribution 
of the various channels even after adjusting the overall normalization 
to data. We determine the relative normalization of the various channels 
by comparing the existing information on the NRQCD matrix 
elements obtained by standard leading order NRQCD analyses. 
The colour singlet matrix element can be computed from the 
wave function at the origin.  The Buchm\"uller-Tye potential is often 
adopted with the result \cite{EQ95}
\begin{equation}
\langle {\cal O}^{J/\psi}_1(^3\!S_1) \rangle
= \frac{9|R(0)|^2}{2\pi} = 1.16 \, \mbox{GeV}^3.
\end{equation}
Due to our particular treatment of the colour singlet contribution as 
described below, we do not need 
this matrix element in $B$ decay. The colour octet 
matrix elements are determined by fits to $J/\psi$ production in a 
variety of production processes.  
$\langle {\cal O}^{J/\psi}_8(^3\!S_1) \rangle$ 
is best determined from $J/\psi$ production in hadron-hadron 
collisions at large transverse momentum \cite{BF95,CL96,BK97}, or,
perhaps, from charmonium production in $Z^0$ decays \cite{BLR99}. 
Given uncertainties from unknown higher order perturbative corrections 
a reasonable range is 
\begin{equation}
\langle {\cal O}^{J/\psi}_8(^3\!S_1) \rangle
= (0.5-1.0)\cdot 10^{-2} \, \mbox{GeV}^3.
\end{equation}
The determination of the other two matrix elements from hadron-hadron 
collisions is much more uncertain. Assuming the above range for 
$\langle {\cal O}^{J/\psi}_8(^3\!S_1) \rangle$ a significant constraint 
on 
\begin{equation}
M_k^{J/\psi}
(^1\!S_0^{(8)},^3\!P_0^{(8)}) = \langle {\cal O}^{J/\psi}_8(^1S_0) \rangle
+ \frac{k}{m_c^2} \langle {\cal O}^{J/\psi}_8(^3P_0) \rangle \,.
\end{equation}
with $k=3.1$ arises from the integrated $J/\psi$ branching in $B$ 
decay itself \cite{BMR99}. A reasonable range is 
\begin{equation}
M_{3.1}^{J/\psi}
(^1\!S_0^{(8)},^3\!P_0^{(8)}) = (1.0-2.0)\cdot 10^{-2} \, \mbox{GeV}^3.
\end{equation}
As default we take the value $M_{3.1}^{J/\psi} =
1.5 \cdot 10^{-2} \,\mbox{GeV}^3$ for $m_c = 1.5\, \mbox{GeV}$  and 
assume that it originates from both parts equally. We then investigate 
the modification of the spectrum when 
$M_{3.1}^{J/\psi}(^1\!S_0^{(8)},^3\!P_0^{(8)}) $ is saturated by only 
one of the two matrix elements and when the relative contribution of 
$\langle {\cal O}^{J/\psi}_8(^3\!S_1) \rangle$ and 
$M_{3.1}^{J/\psi}(^1\!S_0^{(8)},^3\!P_0^{(8)}) $ is varied 
as allowed by the ranges of values given. At the end we discard the 
absolute normalization that would be implied by these values and 
re-fit it to data as already mentioned.

A final comment concerns the treatment of the two-body modes 
$B \to J/\psi K$ and $B \to J/\psi K^*$, which appear as sharp 
resonances in the $J/\psi$ momentum spectrum. Neither the 
ACCMM model nor the shape function for $c\bar{c}$ fragmentation 
applies to these resonance contributions. Fortunately, the information 
provided in \cite{CLEO95} allows us to subtract these contributions 
from the momentum spectrum. We then assume that the two 
resonant contributions are dual to the colour singlet contribution, 
while the rest of the spectrum corresponds to the colour octet 
contribution. This appears plausible, because we expect colour octet 
$c\bar{c}$ pairs to fragment into multi-body final states,  with only a 
small probability that the emitted soft gluons reassemble with the 
spectator quark to form a single hadron. Hence, the experimental 
spectrum shown in the following plots refers to the CLEO data with 
$B \to J/\psi K$ and $B \to J/\psi K^*$ subtracted and it is compared 
with colour octet contributions only. The integrated branching fraction 
from the resonance subtracted spectrum is $0.53 \, \%$. Of course, 
indirect contributions from $B \to \psi' X$ and $B \to \chi_c X$ with 
subsequent decay into $J/\psi$ are also subtracted. 

\begin{figure}
  \vspace*{-1.6cm}
  \begin{center}
    \leavevmode
    \vspace*{-0.4cm}
    \includegraphics[width=.7\textwidth]{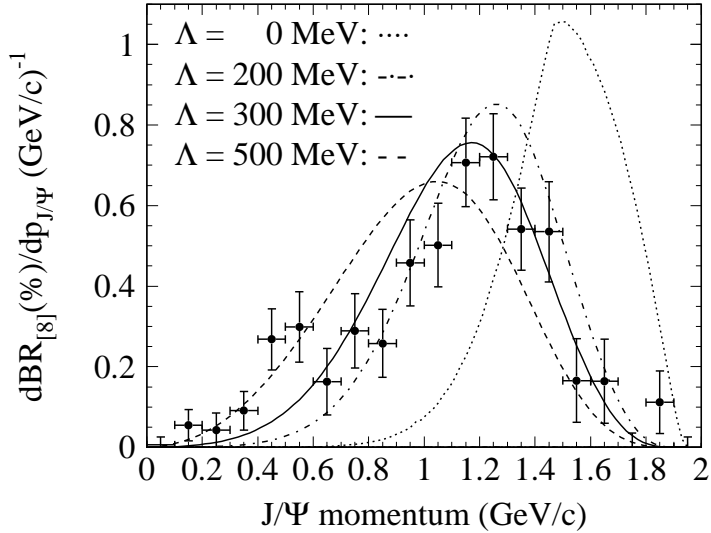}
  \end{center}
  \caption{Sum of colour octet modes $d{\rm BR}_{[8]}/dp_R[n]$ with $n =
    \{^1\!S_0^{(8)}, \, ^3\!P_0^{(8)}, \, ^3\!S_1^{(8)} \}$ to the differential
    branching ratio $d{\rm BR}/dp_R$ of the decay $B \to J/\psi X$ for
    various values of the shape function parameter $\Lambda$. The ACCMM
    model parameters are fixed at $p_F = 300$ GeV and $m_{sp} = 150$ GeV.}
  \label{fitting}
\end{figure}

We have implemented the five-fold integration that leads to the final 
$J/\psi$ momentum spectrum into a Monte Carlo program that uses 
the VEGAS routine described in \cite{VEGAS}. Parameters are 
chosen as follows: $G_F = 1.166 \cdot 10^{-5} \,\mbox{GeV}^{-2}$, 
$|V_{cb}| = 0.039$, $M_\Upsilon = 10.580 \,\mbox{GeV}$, 
$M_B = 5.279 \,\mbox{GeV}$ and $M_\psi =
3.097 \,\mbox{GeV}$. The Wilson coefficient $C_{[8]}(\mu)$ is taken at the
scale $\mu = 4.8$ GeV, which yields $C_{[8]} = 2.19$. The result 
compared to data is shown in Fig.~\ref{fitting} for various values of 
the shape function parameter $\Lambda$ (see the ansatz (\ref{ansatz}) 
and (\ref{cutparam})). Here we have fixed the ACCMM model 
parameters to $p_F = 300$ MeV, motivated by the CLEO analysis of 
semi-leptonic $B$ decay \cite{CLEOpF}, and $m_{sp} =150$ MeV. It is 
clearly seen that the effect of $c\bar{c}$ fragmentation is necessary 
to reproduce the data for this choice of ACCMM parameters. 
Increasing $\Lambda$ shifts the maximum of the spectrum to
lower values of $p_R$. We get the best fit for $\Lambda = 300$ MeV, where
$\chi^2 = 30.2/20$ d.o.f.. 

\begin{figure}
  \vspace*{-1.6cm}
  \begin{center}
    \leavevmode
    \vspace*{-0.4cm}
    \includegraphics[width=.7\textwidth]{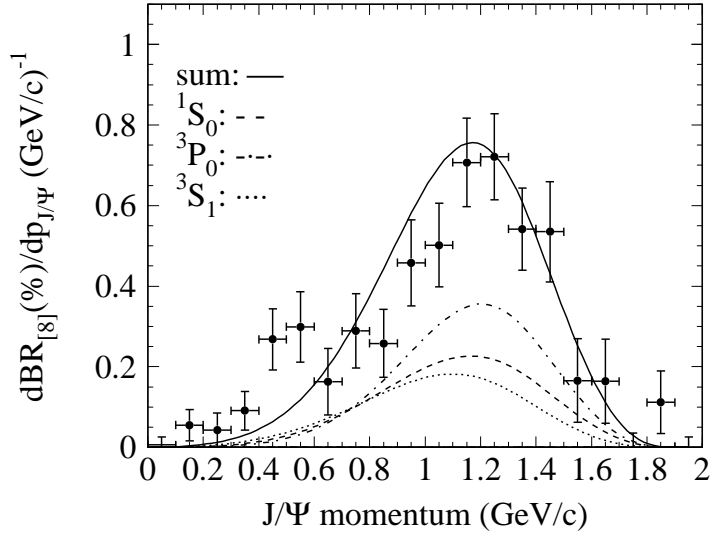}
  \end{center}
  \caption{Contributions of the different colour octet modes $n =
    \{^1\!S_0^{(8)}, \, ^3\!P_J^{(8)}, \, ^3\!S_1^{(8)} \}$ to the sum
    $d{\rm BR}/dp_R$ of the differential branching ratio. The shape
    function and the ACCMM parameters are fixed to $\Lambda = 300$ MeV,
    $p_F = 300$ MeV and $m_{sp} = 150$ MeV.}
  \label{modes}
\end{figure}

\begin{figure}
  \vspace*{-1.6cm}
  \begin{center}
    \leavevmode
    \vspace*{-0.4cm}
    \includegraphics[width=.7\textwidth]{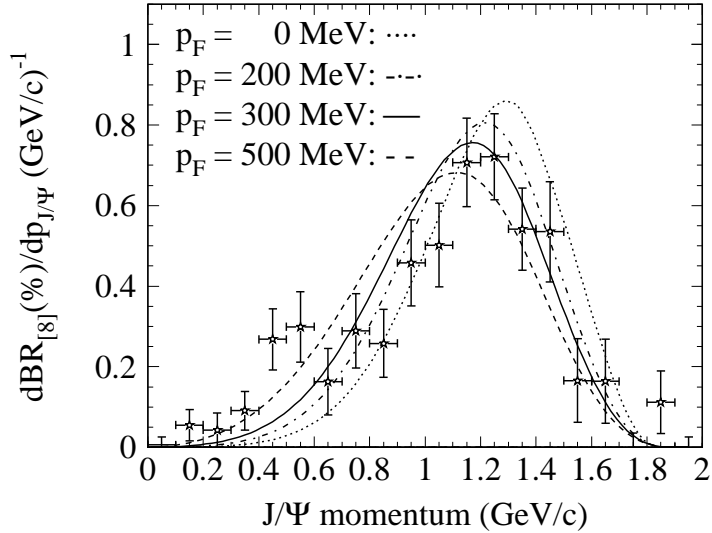}
  \end{center}
  \caption{Sum of colour octet modes $d{\rm BR}_{[8]}/dp_R[n]$ with $n =
    \{^1\!S_0^{(8)}, \, ^3\!P_0^{(8)}, \, ^3\!S_1^{(8)} \}$ to the differential
    branching ratio $d{\rm BR}/dp_R$ of the decay $B \to J/\psi X$ for
    various values of the ACCMM parameter $p_F$. The shape function parameter
    $\Lambda = 300$ MeV and the spectator mass $m_{sp} = 150$ of the ACCMM
    model are kept fixed.}
  \label{pFdep}
\end{figure}

In order to estimate the uncertainty of this fit we investigated the 
sensitivity of the best-fit $\Lambda$ to the variation of the relative 
normalization of the various $c\bar{c}$ production channels as 
described above and to the ACCMM parameter $p_F$. Fig.~\ref{modes} 
shows the best-fit result of Fig.~\ref{fitting} broken down into 
the separate contributions of the three colour octet channels. 
Each channel peaks approximately at the same value $p_R$ and has
similar shapes, although the ${}^3\!S_1$ contribution is somewhat broader 
due to the choice of $c=1.5$ in (\ref{cutparam}). (Varying $c$ between
1 and 2 does not change our fit significantly.) Thus the result of fitting
$\Lambda$ is rather stable under changing the weightings of the different
modes. Both, increasing the relative contribution of $M_{3.1}^{J/\psi}$ and 
saturating it by only one of its matrix elements leads to variations of 
$\Lambda$ of about 50 MeV. There is an obvious anti-correlation between 
the size of $\Lambda$ and of $p_F$, although the effect is not as 
large as one may expect.  Figure \ref{pFdep} shows
the spectra for different values of $p_F$ while $\Lambda$ is fixed to 
300 MeV. We obtain that the spectrum is slightly wider for higher
values of $p_F$. But even for $p_F = 500$ MeV the best-fit $\Lambda$ 
would remain of order 200 MeV.

We conclude from this analysis that the kinematics of soft gluon emission 
has to be accounted for to describe the data on $J/\psi$ momentum spectra 
and that our shape function model provides a satisfactory description 
of the spectrum shape, if the parameter $\Lambda$ is chosen in the 
range
\begin{equation}
\Lambda = 300^{+50}_{-100} \,\mbox{MeV}.
\end{equation}
This result agrees perfectly with the velocity 
scaling rules, which lead to the estimate $\Lambda \sim m_c v^2\sim 
\Lambda_{\rm QCD}$. It is also worth noting that the partonic 
spectrum behind Fig.~\ref{fitting} is a pure delta-function so that 
the smearing due to $c\bar{c}$ fragmentation and Fermi motion extends 
almost over the entire accessible momentum range. Only for rather 
small $J/\psi$ momentum, there would be a visible tail due to perturbative 
hard gluon radiation \cite{BMR99}.

Finally let us comment on the $J/\psi$ momentum spectra in 
\cite{PPS97,AH99} based on the effect of Fermi motion only. (Earlier 
results \cite{BKLP81} were based on the colour singlet model and 
will not be discussed.) These works also report acceptable fits of 
the $J/\psi$ momentum spectrum, however with a larger value of 
$p_F\approx 550\,$MeV, as one may expect when $c\bar{c}$ fragmentation 
effects are neglected. However, even this large value of $p_F$ is 
obtained only, because the $K$ and $K^*$  resonances, 
which sit at large values of 
$p_R$ have been included, even though the ACCMM model 
cannot be applied to them. If these contributions are subtracted, as 
done in the present analysis, a satisfactory fit is not obtained  
with the ACCMM model alone.

\section{Inelastic $J/\psi$ photoproduction}
\label{photoproduction}

\addtocounter{footnote}{-9}

\noindent
In this section we discuss the energy spectrum in inelastic $J/\psi$
photoproduction. This is perhaps the most interesting application of the 
shape function model developed in this paper. The colour octet 
contributions to the energy spectrum have been predicted to increase 
rapidly in the endpoint region, where the $J/\psi$ approaches 
its maximal kinematically allowed energy \cite{CK96,BKV98}. If the colour 
octet matrix elements take the values required to fit the normalization 
of production cross sections in hadron-hadron collisions and in $B$ decay, 
this prediction contradicts the data collected at the HERA collider 
\cite{PHOTO}, which show a rather flat energy distribution. The measured 
distribution can in principle be described 
by colour singlet contributions alone, both at leading order and at 
next-to-leading order \cite{K95} in $\alpha_s$.

Several solutions have been proposed to solve this problem for the 
NRQCD approach to charmonium production: 

(a) The relevant colour octet 
matrix elements are smaller than believed \cite{KK98}. The colour 
octet contributions are always small and the shape of the energy 
spectrum is determined by the colour singlet term.

(b) The partonic cross section is modified by intrinsic transverse 
momentum effects. Within a particular model for these effects 
\cite{SMS98} obtains a reduction of the colour octet cross section, while 
the energy dependence is essentially unmodified.

(c) The NRQCD calculation is unreliable for large $J/\psi$ energies 
because of a breakdown of the non-relativistic expansion \cite{BRW97}. 
Resummation of the expansion as discussed earlier leads to 
folding the partonic cross section with a shape function. It is expected 
that this leads to a depression of the spectrum at large $J/\psi$ 
energies, because some energy is lost for radiation in the fragmentation 
of the colour octet $c\bar{c}$ pair. 

In this section we pursue suggestion (c), which has not been implemented 
in practice yet. Let us note that, irrespective of the issue of 
normalization, this is the only solution that addresses the fact that 
the shape of the colour octet spectrum obtained from a partonic 
calculation is unphysical for large $J/\psi$ energies.

The section is organized as follows. In parallel with the 
discussion of the $B$ decay
we begin with kinematics and by listing the relevant partonic 
subprocesses $\gamma +g \to c\bar{c}[n] + g$.\footnote{Photon-quark 
scattering is a small correction on the scale of effects we are going 
to discuss, and relative to photon-gluon fusion. We omit these subprocesses  
for simplicity.} We then incorporate the fragmentation of the 
$c\bar{c}$ pair via our shape function ansatz and discuss the 
modification of the energy spectrum. For the sake of demonstration, 
we compare the result to HERA data, although we shall see that this 
comparison is problematic from a theoretical point of view.

\subsection{Kinematics of photoproduction}

\noindent
The quantity of interest is $d\sigma/dz$, where
\begin{equation}
z = \frac{p_R \cdot p_p}{p_\gamma \cdot p_p}, 
\end{equation}
and $p_R$, $p_p$ and $p_\gamma$ denote the $J/\psi$, proton and photon 
momentum, respectively. In the proton rest frame $z$ is the fraction of 
the photon energy transferred to the $J/\psi$. In the photon-proton 
centre-of-mass system (cms) we define
\begin{eqnarray}
p_\gamma &=&     \frac{\sqrt{s}}{2} \, (1,0,0,-1)               , \\
p_g      &=& x_g p_p = x_g \frac{\sqrt{s}}{2} \, (1,0,0,+1)     , \\
p_R      &=& (E_R,p_T,0,p_R^z)                                  ,
\end{eqnarray}
where $s = (p_p + p_\gamma)^2$ is the cms energy and $x_g$ the gluon
momentum fraction of the proton momentum. Note that $z$ and $p_T$ 
refer to the physical $J/\psi$ particle. In the present context they 
cannot be identified with the corresponding quantities of the progenitor 
$c\bar{c}$ pair, which we denote by $z_{c\bar{c}}$ and $p_{T,c\bar{c}}$. 
Using $p_R^2 = M_R^2$, we express the $J/\psi$ energy $E_R$ and its 
longitudinal momentum $p_R^z$ in terms of its transverse 
momentum $p_T$ and $z$:
\begin{equation}
E_R = \frac{z^2 s + p_T^2 + M_R^2}{2z \sqrt{s}} , \qquad
p_R^z = - \frac{z^2 s - p_T^2 - M_R^2}{2z \sqrt{s}}.
\end{equation}

The convolution with the shape function, (\ref{master}), requires 
$\alpha$ and $\beta$, defined by (\ref{albe}) in the quarkonium rest
frame.\footnote{Contrary to the previous section we now use ``hats'' 
to denote quantities defined in the $J/\psi$ rest frame. Non-invariant 
quantities without hat refer to the photon-proton cms frame with $z$ axis 
in the direction of the proton momentum.} 
According to our convention, the $\hat{z}$-axis is defined in the 
direction of $-\hat{\bff{P}}_{in}$ with 
$\hat{\bff{P}}_{in} = \hat{\bff{p}}_\gamma +\hat{\bff{p}}_g$. 
Writing 
\begin{eqnarray}
\hat{p}_\gamma & = &
\left( \hat{E}_\gamma, \; \hat{p}_\perp, \; 0, \; \hat{p}_\gamma^z \right) ,
\\
\hat{p}_g & = &
\left( \hat{E}_g, \; - \hat{p}_\perp, \; 0, \; \hat{p}_g^z \right) ,
\\
\hat{P}_{in} & = &
\left( \hat{E}_{in}, \; 0, \; 0, \; \hat{P}_{in}^z \right) ,
\end{eqnarray}
and performing the Lorentz transformation explicitly, we obtain 
\begin{eqnarray}
\hat{E}_\gamma & = &
\frac{M_R^2 + p_T^2}{2 M_R z} ,
\\
\hat{p}_\perp & = &
\frac{p_T z x_g s}{\lambda^{1/2}(M_R^2,-p_T^2,x_g s z^2)} ,
\\
\hat{p}_\gamma^z & = &
- \frac{z^2 x_g s (p_T^2 - M_R^2) + (p_T^2 + M_R^2)^2}
       {2 z M_R \lambda^{1/2}(M_R^2,-p_T^2,x_g s z^2)} ,
\\
\hat{E}_g & = & \frac{z x_g s}{2 M_R} ,
\label{boosted} \\
\hat{p}_g^z & = &
- \frac{z x_g s (z^2 x_g s + p_T^2 - M_R^2)}
       {2 M_R \lambda^{1/2}(M_R^2,-p_T^2,x_g s z^2)} 
\end{eqnarray}
with $\lambda(x,y,z)=x^2+y^2+z^2-2 x y-2 x z-2 y z$, and 
\begin{equation}
\label{inhat}
\hat{E}_{in} =  \frac{M_R^2 + p_T^2 +x_g s z^2}{2 M_R z} ,
\qquad
\hat{P}_{in}^z = 
- \frac{\lambda^{1/2}(M_R^2,-p_T^2,x_g s z^2)}{2 M_R z}.
\end{equation}
The previous line gives $\alpha$ and $\beta$, defined as 
\begin{equation}
\alpha = \hat{E}_{in} + \hat{P}_{in}^z - M_R,
\qquad
\beta = \hat{E}_{in} - \hat{P}_{in}^z - M_R 
\end{equation}
for given $z$ and $p_T$ of the $J/\psi$ in the cms frame.

The Mandelstam variables that appear in the hard production amplitude 
for $\gamma+g\to c\bar{c}[n]+g$ are defined as
\begin{eqnarray}
\hat{s} &=& (\hat{p}_g + \hat{p}_\gamma)^2 = x_g s  , \nonumber    \\
\hat{t} &=& (\hat{p}_{c\bar{c}} - \hat{p}_\gamma)^2 , \label{mandelstam} \\
\hat{u} &=& (\hat{p}_{c\bar{c}} - \hat{p}_g)^2.         \nonumber
\end{eqnarray}
We have to express them in terms of $z$, $p_T$, $x_g$ and 
the energy $\hat{k}_0$ 
and invariant mass $\hat{k^2}$ of the radiated soft partons in the 
$J/\psi$ rest frame.  Recall that  $\hat{p}_{c\bar{c}} 
\equiv \hat{P} + \hat{l} =\hat{p}_R + \hat{k}$ with $\hat{P}=
(2 m_c,\bff{0})$. Hence 
\begin{equation}
\hat{u}=M_{c\bar{c}}(k)^2-\hat{s}-\hat{t},
\end{equation}
where $M_{c\bar{c}}(k)^2=M_R^2+2 M_R\hat{k}_0+\hat{k}^2$ as usual. Next 
parametrize the momentum of the $c\bar{c}$ pair by 
\begin{equation}
\label{pcc}
\hat{p}_{c\bar{c}} = \left(
\hat{E}_{c\bar{c}}, \; \hat{l}_\perp \cos\hat{\phi},
                    \; \hat{l}_\perp \sin\hat{\phi}, \hat{l}_z
                        \right).
\end{equation}
This introduces azimuthal angular 
dependence into the partonic matrix element. This dependence is 
formally small. All $\hat{\phi}$ dependent terms are proportional to 
$\hat{l}_\perp$, and, as discussed in Sect.~\ref{shapelim}, such 
transverse momentum dependence can be neglected in the strict shape 
function limit. In our ansatz, which models the entire spectrum, we also 
have to keep the exact kinematic relations and therefore a non-trivial 
azimuthal average of the hard production amplitude appears in this 
case. With the help of on-shell conditions for the hard emitted gluon we can 
express the components of $\hat{p}_{c\bar{c}}$ by
\begin{eqnarray}
\hat{E}_{c\bar{c}} &=& M_R + \hat{k}_0,
\nonumber \\
\hat{l}_\perp &=& \hat{k}_\perp = 
\frac{[\hat{k}^2 - \alpha (2\hat{k}_0 - \alpha)]^{1/2}
        [\beta (2\hat{k}_0 - \beta) - \hat{k}^2]^{1/2}}{\beta - \alpha} ,
\nonumber \\
\hat{l}_z &=& \hat{k}_z = 
\frac{\hat{k}^2 + \alpha \beta - \hat{k}_0 (\alpha + \beta)}{\beta - \alpha}.
\end{eqnarray}
This result, together with the result for $\hat{p}_\gamma$ and $\alpha$, 
$\beta$, allows us to express $\hat{t}$ in terms of $z$, $p_T$, 
$\hat{k}_0$, $\hat{k}^2$ and $x_g$.

Let us now turn to the hard amplitudes squared of the partonic subprocesses. 
We restrict ourselves to photon-gluon fusion, 
$\gamma + g \to c\bar{c}[n] + g$, where $n$ represents either 
the dominant colour singlet state ${}^3S_1$ or one of the colour 
octet states ${}^1S_0$, ${}^3P_J$, ${}^3S_1$. In terms of Mandelstam 
variables, the spin averaged expressions are \cite{CK96,BKV98,XPYH99}:
\begin{eqnarray}
H[^3S_1^{(1)}](\hat{s},\hat{t},\hat{u},2m_c)
&=& \frac{16 e_c^2 e^2 g_s^2 (2m_c) \left[
        \hat{s}^2 (\hat{t} + \hat{u})^2 + \hat{t}^2 (\hat{u} + \hat{s})^2
        + \hat{u}^2 (\hat{s} + \hat{t})^2 \right]}
        {27 (\hat{s} + \hat{t})^2 (\hat{t} + \hat{u})^2 (\hat{u} + \hat{s})^2},
\\
H[^1S_0^{(8)}](\hat{s},\hat{t},\hat{u},2m_c)
&=& \frac{3 e_c^2 e^2 g_s^2 \hat{s} \hat{u} \left[
 (\hat{s} + \hat{t} + \hat{u})^4 + \hat{s}^4 + \hat{t}^4 + \hat{u}^4 \right]}
 {(2m_c) \hat{t} (\hat{s} + \hat{t})^2 (\hat{t} + \hat{u})^2
 (\hat{u} + \hat{s})^2},
\\
H[^3P_J^{(8)}](\hat{s},\hat{t},\hat{u},2m_c)
&=& \frac{6 e_c^2 e^2 g_s^2}{(2m_c) \hat{t} (\hat{s} + \hat{t})^3
                                (\hat{t} + \hat{u})^3 (\hat{u} + \hat{s})^3}
\nonumber\\
&& \mbox{}
\times \Big[ \hat{t}^6
        (2\hat{s}^3 + 13\hat{s}^2 \hat{u} + 13\hat{s} \hat{u}^2 + 2\hat{u}^3)
\nonumber\\
&& \mbox{}
+ \hat{t}^5
        (4\hat{s}^4 + 47\hat{s}^3 \hat{u} + 70\hat{s}^2 \hat{u}^2
                                        + 47\hat{s} \hat{u}^3 + 4\hat{u}^4)
\nonumber\\
&& \mbox{}
+ \hat{t}^4
        (2\hat{s}^5 + 63\hat{s}^4 \hat{u} + 136\hat{s}^3 \hat{u}^2
                + 136\hat{s}^2 \hat{u}^3 + 63\hat{s} \hat{u}^4 + 2\hat{u}^5)
\nonumber\\
&& \mbox{}
+ \hat{s} \hat{t}^3 \hat{u}
        (47\hat{s}^4 + 132\hat{s}^3 \hat{u} + 190\hat{s}^2 \hat{u}^2
                                        + 132\hat{s} \hat{u}^3 + 47\hat{u}^4 )
\nonumber\\
&& \mbox{}
+ \hat{s} \hat{t}^2 \hat{u}
        (25\hat{s}^5 + 88\hat{s}^4 \hat{u} + 156\hat{s}^3 \hat{u}^2
                + 156\hat{s}^2 \hat{u}^3 + 88\hat{s} \hat{u}^4 + 25\hat{u}^5)
\nonumber\\
&& \mbox{}
+ \hat{s} \hat{t} \hat{u}
        (7\hat{s}^6 + 38\hat{s}^5 \hat{u} + 78\hat{s}^4 \hat{u}^2 + 98\hat{s}^3
        \hat{u}^3 + 78\hat{s}^2 \hat{u}^4 + 38\hat{s} \hat{u}^5 + 7\hat{u}^6)
\nonumber\\
&& \mbox{}
+ 7 \hat{s}^2 \hat{u}^2 (\hat{s} + \hat{u})
                                (\hat{s}^2 + \hat{s} \hat{u} + \hat{u}^2)^2
\Big],
\\
H[^3S_1^{(8)}](\hat{s},\hat{t},\hat{u},2m_c)
&=& \frac{15}{8} H[^3S_1^{(1)}](\hat{s},\hat{t},\hat{u},2m_c).
\end{eqnarray}
Here $e$ is the electromagnetic coupling, $g_s$ the strong coupling and 
$e_c=2/3$ the electric charge of the charm quark.
Note that the NRQCD elements are not part of the hard 
cross sections, but included in the normalization of the radiation 
function $\Phi_n(k)$, see (\ref{normalization}). In case of the 
$P$ wave contribution, the normalization refers to $\langle {\cal O}_8
({}^3P_0)\rangle/m_c^2$ and the corresponding factor $1/m_c^2$ is 
also extracted from $H[^3P_J^{(8)}](\hat{s},\hat{t},\hat{u},2m_c)$. 

The hard amplitudes squared are then expressed as functions of 
$z$, $p_T$, $x_g$, $\hat{k}_0$, $k^2$ and $\hat{\phi}$ and the 
average over the azimuthal angle $\hat{\phi}$ according to 
(\ref{azav}) is performed. The double differential cross section 
for $\gamma + g \to J/\psi + X$ is then given by 
\begin{eqnarray}
\frac{d^2\sigma_{\gamma g}}{dp_T^2 dz} &=& \frac{1}{16\pi^2 z} \, \sum_n
\int\limits_0^{\alpha\beta} \frac{d\hat{k}^2}{2\pi} \!
\int\limits_{(\alpha^2+\hat{k}^2)/(2\alpha)}^{(\beta^2+\hat{k}^2)/(2\beta)}
\!\!\! d\hat{k}_0 \,\, 
\nonumber\\
&&\hspace*{1cm}
\cdot\,\frac{1}{2 \hat{s}} \, \bar{H}_n(z, p_T^2, x_g,\hat{k}_0,\hat{k}^2) \,
\frac{4\pi M_R z}{\lambda^{1/2}(M_R^2,-p_T^2,x_g s z^2)} \, 
\Phi_n(\hat{k}).
\end{eqnarray}
The sum runs over the four $c\bar{c}$ states listed above. Note, 
however, that no shape function is required for the colour singlet 
contribution, since the dominant contribution to the colour singlet 
matrix element does not require emission of soft gluons. For the 
colour singlet contribution we therefore use the ordinary differential 
cross section on the parton level. The final result is obtained by 
folding in the gluon distribution in the proton, $g(x_g,\mu_F)$, and 
integrating over transverse momentum:
\begin{equation}
\frac{d\sigma_{\gamma p}}{dz} =
\int\limits_{p_{T, {\rm min}}^2}^{p_{T, {\rm max}}^2} \!\!\!\! dp_T^2
\int\limits_{x_{g, {\rm min}}}^{1} \!\!\!\! dx_g \, g(x_g,\mu_F) \,
\frac{d^2\sigma_{\gamma g}}{dp_T^2 dz} \,.
\end{equation}
The lower integration limit for $p_T^2$ usually is set by an experimental 
cut. In the present framework such a cut is needed to eliminate the 
contribution from the $2\to 1$ process $\gamma+g\to c\bar{c}[n]$, 
smeared out over a finite range in $p_T$ and $z$ by soft gluon 
emission in the fragmentation of the $c\bar{c}$ pair, and also from 
the initial state. The other bounds are given by
\begin{eqnarray}
p_{T, {\rm max}}^2 & = & (1 - z)(s z - M_R^2) , 
\\
x_{g, {\rm min}} & = & \frac{M_R^2 (1 - z) + p_T^2}{s z (1 - z)} .
\end{eqnarray}
The minimum $p_T$ cut implies that $z<1-p_{T,\rm min}^2/s+\ldots$. 
For large cms energy, as at the HERA collider, this is not a severe 
restriction on the $z$ spectrum.

\subsection{Discussion of the energy spectrum}

\noindent
The following results for the energy spectrum are obtained with the 
GRV94 LO gluon density \cite{GRV} and factorization scale $\mu_F=M_R$, 
where $M_R$ is the $J/\psi$ mass. We also use $\Lambda_{\rm QCD}^{n_f=4}=
0.2$ GeV (consistent with GRV) and  $\alpha_s(M_R)=
0.275$. The cms energy is fixed to an average photon-proton 
cms energy at HERA, $\sqrt{s}=100\,$GeV. We also choose $m_c=1.5\,$ 
GeV for the colour-singlet process.

\begin{figure}[p]
\vspace*{-2cm}
  \begin{center}
  \leavevmode
  \includegraphics[height=0.5\textheight]{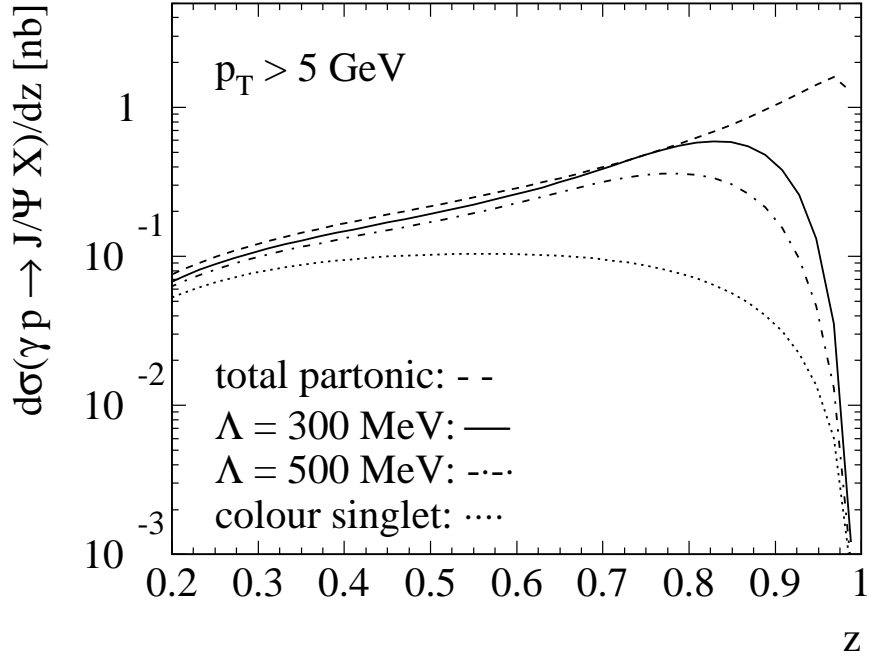}
  \end{center}
\vspace*{-2.5cm}
  \begin{center}
  \leavevmode
  \includegraphics[height=0.5\textheight]{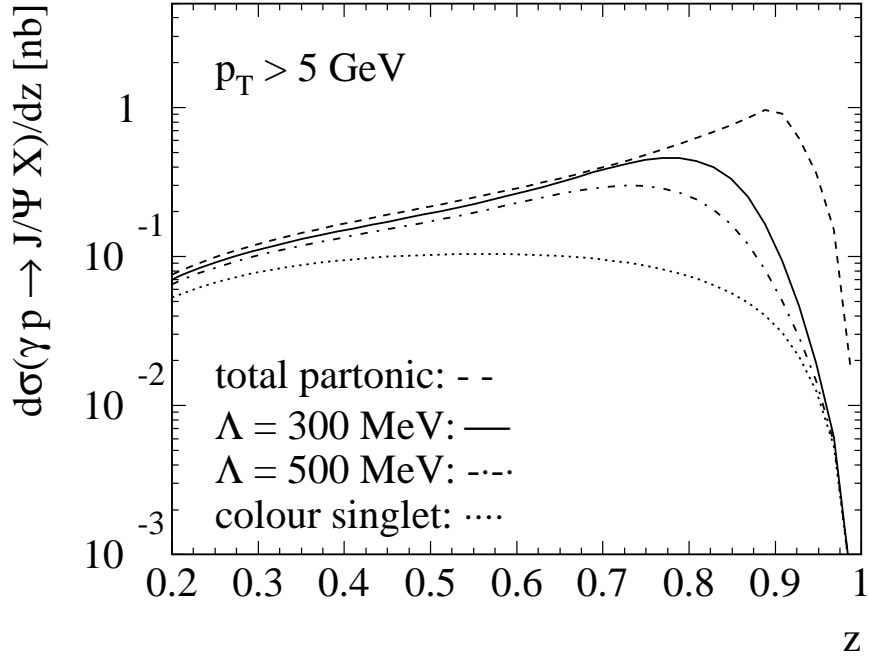}
  \end{center}
\caption[dummy]{\label{fig8}\small
The $J/\psi$ energy spectrum for $\sqrt{s}=100\,$GeV and with a 
transverse momentum cut $p_{T,\rm min} = 5$ GeV. Upper panel: spectrum 
for three values of the shape function parameter $\Lambda =
0\,\mbox{(``total partonic'')}, 300, 500\,$MeV. Dotted: colour singlet 
contribution alone. Lower panel: as upper panel but with   
the ``modified matrix element'' discussed in the text.}
\end{figure}

In Fig.~\ref{fig8} we display the $J/\psi$ energy spectrum 
$d\sigma/d z$ with the $J/\psi$ transverse momentum larger than $5\,$ 
GeV. This cut is larger than the one currently used by the HERA 
experiments. However, it allows us to discuss the effect of 
$c\bar{c}$ pair fragmentation in a situation that is theoretically 
under better control. The curves in the upper plot of Fig.~\ref{fig8} 
show, as expected, that the spectrum turns over and approaches zero 
as $z\to 1$, once some fraction of the photon energy is lost to 
radiation in the fragmentation of the $c\bar{c}$ pair. This turn-over 
occurs at smaller $z$ for larger values of the parameter $\Lambda$, 
which is related to the typical energy lost to radiation in the 
$J/\psi$ rest frame. For $J/\psi$ production in $B$ decay, we 
found that $\Lambda\approx 300\,$MeV fitted the spectrum well. 
Assuming universality of the shape function, this is our preferred 
choice for photoproduction. For comparison, we also display the 
result with $\Lambda=500\,$MeV. Note that these numbers refer to 
the $J/\psi$ rest frame. In another frame, such as the laboratory 
frame, the energy lost to ``soft'' radiation may be large, of 
order $E_R \Lambda/M_R$, where $E_R$ is the $J/\psi$ energy in 
that frame. 

The overall normalization in Fig.~\ref{fig8} and the subsequent figure 
requires comment. The NRQCD matrix elements are chosen as in 
Sect.~\ref{finalresult} on $B$ decay. As in that case the normalization 
has then to be re-adjusted to account for the fact that the effective 
charm quark mass in the hard scattering amplitude is much larger 
than $m_c=1.5\,$GeV, conventionally assumed in fits of NRQCD matrix 
elements. We proceed as follows: The curves labelled ``partonic'' 
(total and colour singlet alone) use $m_c=1.5\,$GeV to allow comparison 
with earlier results. For given $\Lambda$, and for each colour 
octet channel separately, we determine $m_c^{\rm eff}$ defined in 
(\ref{meff}). We then recalculate the partonic curve with $m_c=m_c^{\rm eff}$ 
and determine a normalization ratio by dividing the result 
for $1.5\,$GeV by the second one in the region of $z\approx 0.1-0.4$. 
Finally, we compute the curve including the shape function with the 
given value of $\Lambda$, multiply it 
by this ratio and compare it to the partonic curve for the 
conventional choice $m_c=1.5\,$GeV. The low $z$ region is chosen 
to compute the normalization ratio, since the shape function should 
have little effect on the spectrum far away from the endpoint. As a 
consequence of this procedure the partonic result and the results 
for non-zero $\Lambda$ nearly coincide for small $z$. The normalization 
adjustment is quite large, which reflects the strong $m_c$ dependence 
of the partonic cross sections.

Closer inspection of the upper panel of Fig.~\ref{fig8} shows that 
the spectrum for non-zero $\Lambda$ increases faster for moderate $z$ 
than the partonic spectrum. To understand this effect, we reconsider 
the hard amplitudes squared for the production of a colour octet 
$c\bar{c}$ pair in a ${}^1\!S_0$ or a ${}^3\!P_J$ state as functions 
of $z_{c\bar{c}}$ and $p_{T,c\bar{c}}$. For any fixed $p_{T,c\bar{c}}$ 
the hard amplitudes squared increase as $1/(1-z_{c\bar{c}})^2$ as 
$z_{c\bar{c}}\to 1$, as follows from $\hat{s}=-\hat{t}/(1-z_{c\bar{c}})$ 
and $\hat{u}\approx -\hat{s}$ as $z_{c\bar{c}}\to 1$. 
This causes the troubling increase of colour-octet 
contributions in the partonic calculation.
Now, for any given $z$, 
\begin{equation}
z_{c\bar{c}} = \frac{p_{c\bar{c}} \cdot p_p}{p_\gamma \cdot p_p} \geq 
z 
\end{equation}
as can be seen by going to the proton rest frame. Hence, for fixed $z$, 
 the hard amplitude 
squared is evaluated at larger $z_{c\bar{c}}$, when $\Lambda$ is 
non-zero compared to the partonic result for which $z_{c\bar{c}}=z$. 
Due to the above-mentioned behaviour of the amplitude, sampling the hard 
cross section at larger $z_{c\bar{c}}$  
increases the spectrum. Likewise, the transverse momentum of the 
$c\bar{c}$ pair with respect to the beam axis 
\begin{equation}
p^2_{T, \, c\bar{c}} = (1 - z_{c\bar{c}})
(x_g s z_{c\bar{c}} - M_{c\bar{c}}^2)
\end{equation}
differs from $p_T^2$. This happens for two reasons: first, the loss of 
energy to radiation also implies a loss of transverse momentum with 
respect to the beam axis, if the $J/\psi$ is not parallel to the 
beam axis. Second, the $J/\psi$ can gain transverse momentum by 
recoil against the soft gluons radiated during the fragmentation 
process. For fixed $p_T$, this is preferred to losing transverse momentum, 
because the production amplitude for the $c\bar{c}$ pair increases 
with smaller $p_{T,c\bar{c}}$. The dominant effect is the one 
related to $z_{c\bar{c}}\geq z$. The corresponding increase of the 
spectrum (for $\Lambda$ non-zero and moderate $z$) relative to the 
partonic spectrum is stronger as $p_{T,\rm min}$ is chosen smaller, 
since the hard cross section rises faster for smaller 
$p_{T,\rm min}$ (and would approach the collinear and soft divergence 
at $z=1$, if $p_{T,\rm min}=0$). Finally, at very large $z$, the suppression 
due to the radiation function 
$\Phi_n(k)$ wins and turns the spectrum over to zero.

To illustrate these remarks we define an ad hoc modification of 
the hard cross sections $H_n(z_{c\bar{c}},p_{T,c\bar{c}})$:
\begin{equation}
\label{adhoc}
H_n^{\rm mod}(z_{c\bar{c}},p_{T,c\bar{c}}) = \left\{
\begin{array}{ll}
H_n(0.9,p_{T,c\bar{c}}) & \hspace*{0.3cm}
\mbox{if } z_{c\bar{c}}>0.9,\, p_{T,c\bar{c}} > 1\,\mbox{GeV}
\\ 
H_n(z_{c\bar{c}},1\,\mbox{GeV}) & \hspace*{0.3cm}
\mbox{if } z_{c\bar{c}}<0.9, \,p_{T,c\bar{c}} < 1\, 
\mbox{GeV} \\
H_n(0.9,1\,\mbox{GeV}) & \hspace*{0.3cm}
\mbox{if } z_{c\bar{c}}>0.9, \,p_{T,c\bar{c}} < 1\,\mbox{GeV}
\\
H_n(z_{c\bar{c}},p_{T,c\bar{c}}) &\hspace*{0.3cm}\mbox{else}
\end{array}
\right.
\end{equation}
The energy spectra analogous to the upper panel of Fig.~\ref{fig8} 
but with hard cross sections modified in this way 
are shown in the lower panel 
of this figure. The partonic cross section is modified only for 
$z>0.9$ by construction. The spectra for non-zero $\Lambda$ are 
reduced already at smaller $z$, which shows the sensitivity to 
$z_{c\bar{c}}>0.9$ at such small $z$. 

We emphasize that 
no physical significance should be attached to the lower panel 
of Fig.~\ref{fig8}. The growth of the colour octet cross sections 
at large $z$ is physical and reflects the growth of $2\to 2$ cross 
sections at large rapidity difference due to $t$-channel 
gluon exchange. In the endpoint region $\hat{t}\sim -p_{T,\rm min}^2$ and 
$\hat{s}\sim -\hat{u}\sim p_{T,\rm min}^2/(1-z)$ so that 
$\hat{s}\gg |\hat{t}|$. Higher order corrections to the 
spectrum would involve logarithms of $\hat{s}/(-\hat{t})$. Summation 
of these logarithms with the BFKL (Balitsky-Fadin-Kuraev-Lipatov) 
equation increases the parton cross section in the endpoint region.

\begin{figure}[p]
\vspace*{-2cm}
  \begin{center}
  \leavevmode
  \includegraphics[height=0.5\textheight]{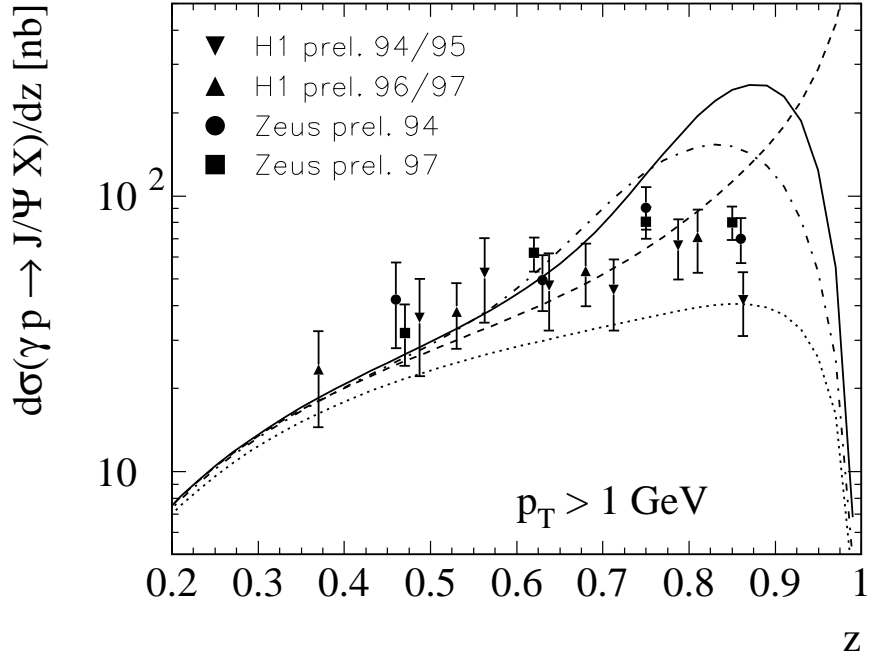}
  \end{center}
\vspace*{-2.5cm}
  \begin{center}
  \leavevmode
  \includegraphics[height=0.5\textheight]{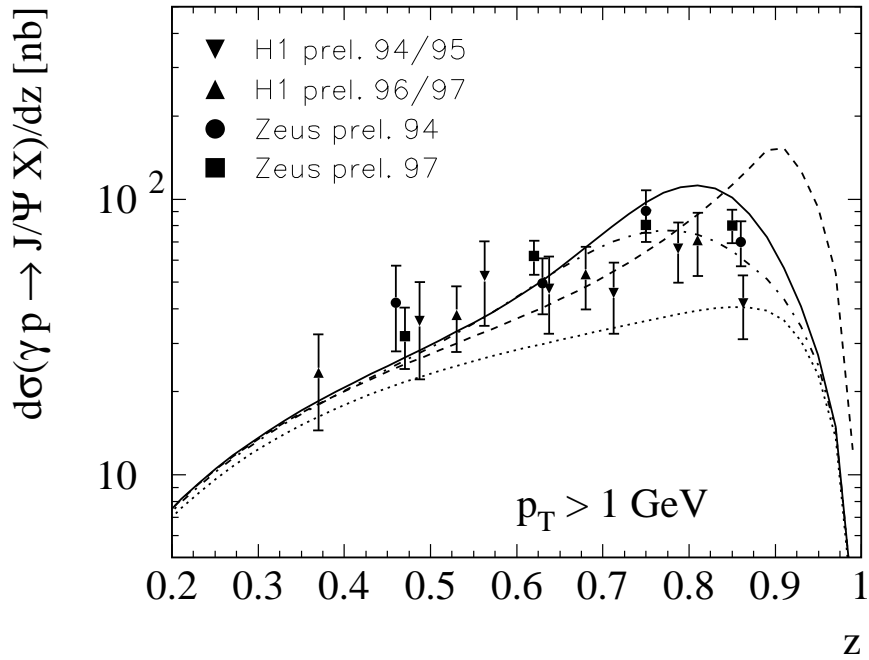}
  \end{center}
\caption[dummy]{\label{fig9}\small
The $J/\psi$ energy spectrum at $\sqrt{s}=100\,$GeV and with 
$p_T>1\,$GeV compared to HERA data \cite{PHOTO}. Upper and lower 
panel as in Fig.~\ref{fig8}. Solid (dash-dotted, dashed) lines 
refer to $\Lambda = 300\, (500, 0)\,$MeV. Dotted: colour singlet 
spectrum alone.} 
\end{figure}

After this discussion for large transverse momentum of the $J/\psi$, 
we display the result for the energy spectrum with the additional 
requirement $p_T>1\,$GeV, which we compare to data from the H1 and ZEUS 
collaborations \cite{PHOTO}. Fig.~\ref{fig9} shows again the conventional 
partonic calculation compared to the calculation with two values of 
$\Lambda$. The lower panel refers to the ad hoc modification of the 
hard cross sections according to (\ref{adhoc}).

The qualitative features evident in the upper panel follow from the 
previous discussion. At large $z$ the spectrum turns over, but at 
intermediate $z$, including the entire region where data exist, 
there is a large enhancement, which follows from the fact that the 
partonic matrix element is sampled 
very close to $z_{c\bar{c}}=1$. Taken at face value, the disagreement 
with data is worse after accounting for $c\bar{c}$ fragmentation 
effects. However, the theoretical prediction with small transverse 
momentum cut is unreliable at large $z$. With no $p_T$ cut at all, 
we expect that the $z$ spectrum is drastically modified at large $z$ 
after accounting for the $2\to 1$ process, the virtual corrections to it, 
and soft gluon radiation from the initial gluon. Owing to the 
sensitivity to large $z_{c\bar{c}}$, the theoretical prediction is 
more sensitive to these modifications when gluon radiation in 
$c\bar{c}$ fragmentation is included. An indication of this is 
provided by plotting the spectrum with the modified partonic cross 
section. This modification of the partonic cross section, although ad hoc, 
may give a qualitative representation of the effects to be expected 
from soft gluon resummation. The lower panel of Fig.~\ref{fig9} shows 
that the unphysical enhancement is largely reduced in this case, although 
it does not disappear completely. If reality turned out to resemble 
the lower panel, it would be difficult to disentangle colour octet 
contributions, given the additional normalization uncertainties of 
both, the colour singlet and the colour octet contributions. In this 
case a $J/\psi$ polarization measurement would provide useful 
additional information \cite{BKV98}.

The results of this analysis can be summarized as follows: with the 
small transverse momentum cut on the $J/\psi$ currently used by 
both HERA collaborations, the region $z>0.7$ is beyond theoretical 
control. This remains true even after resummation of large higher 
order NRQCD corrections via the shape function, since the hard 
partonic cross section is sensitive to other modifications that are 
also difficult to control theoretically at such small transverse 
momentum. At present, the experimental data cannot be interpreted as 
providing evidence for or against the presence of colour octet 
contributions in photoproduction. It is not 
necessary to reduce the colour octet matrix elements as suggested 
in \cite{KK98} to arrive at this conclusion. This is welcome as 
matrix elements of the order quoted in Sect.~\ref{bsect} seem to be needed 
to account for the observed branching fraction of $B\to J/\psi X$. 

The situation in photoproduction remains unsatisfactory. In our 
opinion, nothing is learnt on quarkonium production mechanisms, 
if a small transverse momentum cut is used. We therefore recommend 
that future increases in integrated luminosity should not 
be used to reduce the experimental errors on the present analysis, 
but to increase the transverse momentum cut at the expense of 
statistics.
 
\section{Conclusion}
\label{conclusion}

\noindent 
In this paper we provided a first investigation of the kinematic effect 
of soft emission in the fragmentation process $c\bar{c}[n]\to J/\psi + 
X$. In the NRQCD factorization approach to inclusive quarkonium 
production these effects appear as kinematically enhanced higher order 
corrections in the NRQCD expansion \cite{MW97,BRW97}, which 
become important near the upper endpoint of quarkonium energy/momentum 
spectra. The shape function formalism discussed in \cite{MW97,BRW97} 
resums these corrections and allows us to extend to validity of 
the NRQCD approach closer to the endpoint, although the entire spectrum 
is not covered even after this resummation. In the present paper, we 
implemented the kinematics of soft gluon radiation exactly and used 
an ansatz for the probability of radiation of soft gluons. This allows 
us to cover the entire energy spectrum, although in a model-dependent 
way. The model is consistent with the NRQCD shape function formalism 
in the energy region where the later applies. This situation is 
similar to the relation of the ACCMM model to the heavy quark 
expansion in inclusive semi-leptonic decays of $B$ mesons. The main 
result is provided by (\ref{master}), which applies to a general 
inclusive quarkonium production process, when the partonic final 
state before fragmentation consists of a $c\bar{c}$ pair and 
one additional massless, energetic parton.

We then proceeded to two 
applications of the formalism. These applications 
are not necessarily the simplest ones conceivable, but they seem to 
be most interesting. We first considered $J/\psi$ production in $B$ 
decay, which proceeds through colour octet states by a large fraction. 
In this case the effect of emission in fragmentation of colour 
octet $c\bar{c}$ pairs has to be disentangled from Fermi motion of 
the $b$ quark in the $B$ meson. We found that the description of the 
spectrum improves significantly, when soft radiation is included, 
and if the parameter $\Lambda$ that controls the energy scale for soft 
radiation is chosen around $300\,$MeV. The shape function defined 
in \cite{BRW97} is production process independent. Assuming universality 
of our ansatz over the entire energy range, the same ansatz is used for 
inelastic $J/\psi$ photoproduction.  We found that the energy spectrum 
turns over at $z\approx 0.8-0.9$, to be compared with the partonic 
spectrum that rises towards $z=1$. However, at $z<0.8$, the colour 
octet contributions to the spectrum still grow faster than the colour 
singlet contribution. Due to the increase of the partonic cross section, 
the increase is in fact faster in this intermediate $z$ region 
after $c\bar{c}$ fragmentation effects are included. We also concluded 
that the transverse momentum 
cut $p_T>1\,$GeV, presently used by the HERA experiments, 
is too small to arrive at a reliable theoretical prediction. Hence, 
no conclusion regarding colour octet contributions and the 
validity of the NRQCD formalism can presently be drawn from HERA 
data.

The formalism developed in this paper could be applied to 
other production processes, in which the $J/\psi$ energy is 
measured. Another interesting extension is quarkonium decays, 
when the energy of one of the decay particles is measured, 
such as the photon energy in $\eta_c\to \gamma+X$ and 
$J/\psi\to \gamma+X$. Since decay processes are less affected by  
theoretical uncertainties related to colour recombination and 
initial state radiation than production processes, this may lead 
to theoretically better controlled applications of the shape function 
formalism.

\vspace*{0.5cm}
\noindent
{\em Acknowledgements.}  
We would like to thank M.~Kr\"amer, T.~Mannel and I.Z. Rothstein 
for useful comments.
S.W. thanks the CERN Theory Group for the hospitality on several occasions.
This work was supported in part by the EU Fourth Framework Programme 
`Training and Mobility of Researchers', Network `Quantum Chromodynamics and 
the Deep Structure of Elementary Particles', contract FMRX-CT98-0194 
(DG 12 - MIHT), by the Landesgraduiertenf\"orderung of the state 
Baden-W\"urttemberg, and by the Graduiertenkolleg ``Elementarteilchenphysik 
an Beschleunigern''. S.W. is part of the DFG-Forschergruppe 
Quantenfeldtheorie, Computeralgebra und Monte-Carlo-Simulation.


\end{document}